\documentclass[twocolumn]{aastex63}

\usepackage{CJKutf8}

\usepackage{amsmath}
\usepackage{amssymb}
\usepackage{amsfonts}
\usepackage{graphicx}
\usepackage{wasysym}
\usepackage{multirow}
\usepackage{mdframed}
\usepackage[T1]{fontenc}

\hypersetup{linkcolor=blue,citecolor=blue,filecolor=cyan,urlcolor=blue}

\newcommand{\water}{\mbox{H$_2$O}}

\newcommand{\hcop}{\mbox{HCO$^+$}}

\newcommand{\kms}{\mbox{km\,s$^{-1}$}}

\newcommand{\cc}{\mbox{cm$^{-3}$}}

\newcommand{\msol}{\mbox{$M_\odot$}}
\newcommand{\msolpyr}{\mbox{$M_\odot$\,yr$^{-1}$}}
\newcommand{\mujypbm}{\mbox{$\mu$Jy\,beam$^{-1}$}}
\newcommand{\mjypbm}{\mbox{mJy\,beam$^{-1}$}}

\newcommand{\hii}{\mbox{H\,{\sc ii}}}

\received{- -, --}
\revised{- -, --}
\accepted{- -, --}
\submitjournal{ApJ Letters}

\shorttitle{Massive Cluster Formation in the Central Molecular Zone}
\shortauthors{Lu et al.}

\begin{document}
\begin{CJK}{UTF8}{gbsn}

\title{ALMA Observations of Massive Clouds in the Central Molecular Zone: Jeans Fragmentation and Cluster Formation}
\correspondingauthor{Xing Lu}
\email{xinglv.nju@gmail.com, xing.lu@nao.ac.jp}

\author[0000-0003-2619-9305]{Xing Lu (吕行)}
\affiliation{National Astronomical Observatory of Japan, 2-21-1 Osawa, Mitaka, Tokyo, 181-8588, Japan}

\author{Yu Cheng (程宇)}
\affiliation{Department of Astronomy, University of Virginia, Charlottesville, VA 22904, USA}

\author{Adam Ginsburg}
\affiliation{Department of Astronomy, University of Florida, Gainesville, FL 32611, USA}

\author{Steven N.\ Longmore}
\affiliation{Astrophysics Research Institute, Liverpool John Moores University, 146 Brownlow Hill, Liverpool L3 5RF, UK}

\author{J.~M.~Diederik Kruijssen}
\affiliation{Astronomisches Rechen-Institut, Zentrum f\"{u}r Astronomie der Universit\"{a}t Heidelberg, M\"{o}nchhofstra\ss e 12-14, D-69120 Heidelberg, Germany}

\author{Cara Battersby}
\affiliation{University of Connecticut, Department of Physics, 196 Auditorium Road, Unit 3046, Storrs, CT 06269, USA}

\author{Qizhou Zhang}
\affiliation{Center for Astrophysics | Harvard \& Smithsonian, 60 Garden Street, Cambridge, MA 02138, USA}

\author{Daniel L.\ Walker}
\affiliation{National Astronomical Observatory of Japan, 2-21-1 Osawa, Mitaka, Tokyo, 181-8588, Japan}
\affiliation{Joint ALMA Observatory, Alonso de C\'{o}rdova 3107, Vitacura 763 0355, Santiago, Chile}

\begin{abstract} % =<250 words
We report Atacama Large Millimeter/submillimeter Array (ALMA) Band 6 continuum observations of 2000~AU resolution toward four massive molecular clouds in the Central Molecular Zone of the Galaxy. To study gas fragmentation, we use the dendrogram method to identify cores as traced by the dust continuum emission. The four clouds exhibit different fragmentation states at the observed resolution despite having similar masses at the cloud scale ($\sim$1--5~pc). Assuming a constant dust temperature of 20~K, we construct core mass functions of the clouds and find a slightly top-heavy shape as compared to the canonical initial mass function, but we note several significant uncertainties that may affect this result. The characteristic spatial separation between the cores as identified by the minimum spanning tree method, $\sim$$10^4$~AU, and the characteristic core mass, 1--7~\msol{}, are consistent with predictions of thermal Jeans fragmentation.  The three clouds showing fragmentation may be forming OB associations (stellar mass $\sim$10$^3$~\msol{}). None of the four clouds under investigation seem to be currently able to form massive star clusters like the Arches and the Quintuplet ($\gtrsim$10$^4$~\msol{}), but they may form such clusters by further gas accretion onto the cores.
\end{abstract}


%%%%%%%%%%%%%%%%%%%%%
\section{INTRODUCTION}\label{sec:intro}
The Central Molecular Zone (CMZ), the inner $\sim$$500$~pc of our Galaxy, contains more than $10^7$~\msol{} of molecular gas with intriguing star formation properties \citep{morris1996,longmore2013a}. On the one hand, young massive star clusters, including the Arches and the Quintuplet with $\sim$$10^4$~\msol{} stellar masses and about 100 O-type stars, are found in the CMZ \citep{figer1999b,luj2018} and are suggested to have formed in-situ a few Myr ago \citep{stolte2014,kruijssen2015}. On the other hand, the current star formation in the CMZ is measured to be about 10 times less efficient than that in the Galactic disk \citep{longmore2013a,kruijssen2014,barnes2017}. It is then a question whether any molecular clouds in the CMZ have the potential of forming Arches/Quintuplet-like clusters. Based on the detection of $\sim$60 massive young stellar objects in the Sgr~B2(M) region, one of the most actively star forming sites in the CMZ, \citet{ginsburg2018a} estimated the total stellar mass to be $\sim$$10^4$~\msol{}, making it a likely progenitor of massive clusters.

Are the other clouds in the CMZ forming massive clusters? Outside of Sgr~B2, only a number of clouds in the CMZ have been found to actively form high-mass ($>$8~\msol{}) stars \citep{kauffmann2017a,lu2019a,lu2019b}. In particular, our recent observations of ultracompact (UC) \hii{} regions and masers reveal the 20~\kms{} cloud, the 50~\kms{} cloud, Sgr~B1-off (also known as Dust Ridge clouds e/f), and Sgr C as prominent high-mass star forming clouds \citep{lu2019a,lu2019b}. Previous observations targeting relatively evolved phases of star formation in these clouds (e.g., \hii{} regions, infrared emission from young stellar objects) suggest inefficient star formation, and none of them seem to be forming Arches/Quintuplet-like clusters \citep{mills2011,immer2012b,walker2018}.

Could there exist a population of very early phase star formation that is still deeply embedded in the four clouds, but has been missed previously? Such incipient star formation can be invisible in free-free or infrared emission or masers, but may be revealed by gas fragmentation that leads to prestellar cores. To investigate this possibility, we observe the four clouds using the Atacama Large Millimeter/submillimeter Array (ALMA) at high angular resolutions. Here we report the continuum observations and discuss the implications to star formation in these clouds. Throughout this Letter we adopt a distance of 8.178~kpc to the CMZ \citep{gravity2019}.

%%%%%%%%%%%%%%%%%%%%%
\section{OBSERVATIONS AND DATA REDUCTION}\label{sec:obs}
\begin{figure*}[p]
\centering
\begin{tabular}{@{}p{0.45\textwidth}@{}p{0.55\textwidth}@{}}
\begin{tabular}[c]{@{}c@{}}
\includegraphics[width=0.45\textwidth]{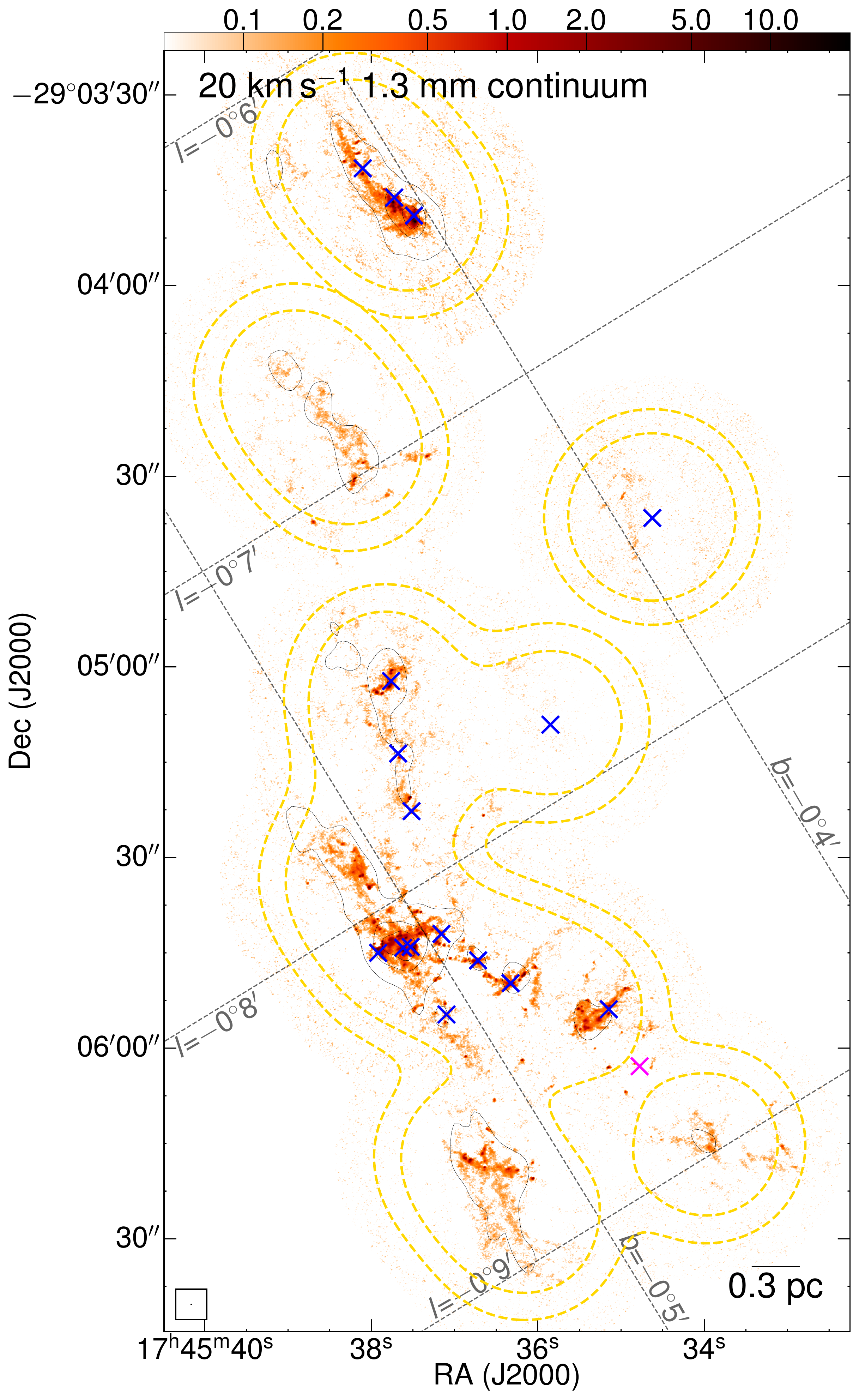} \\
\includegraphics[width=0.35\textwidth]{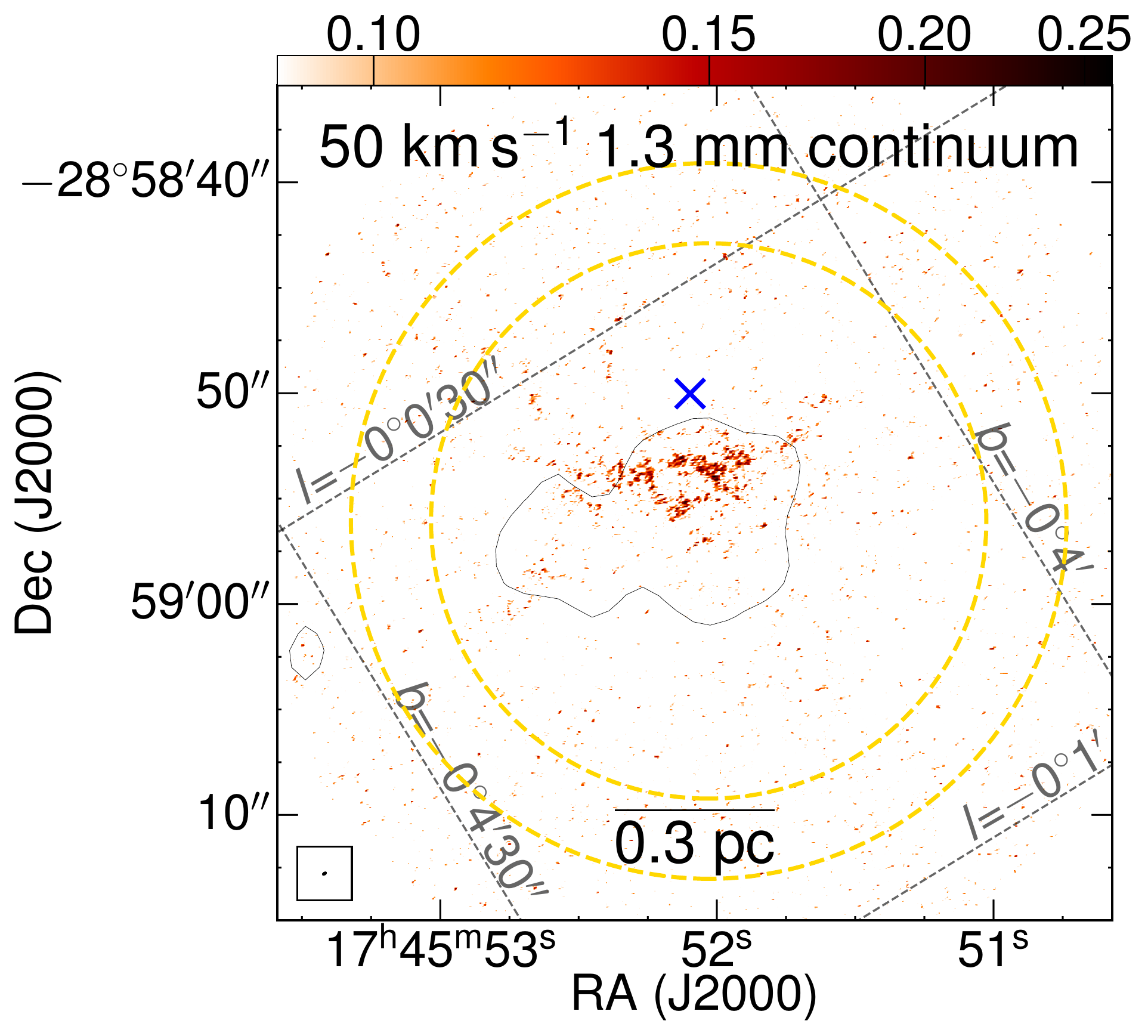}
\end{tabular}
&
\begin{tabular}[c]{@{}c@{}}
\includegraphics[width=0.45\textwidth]{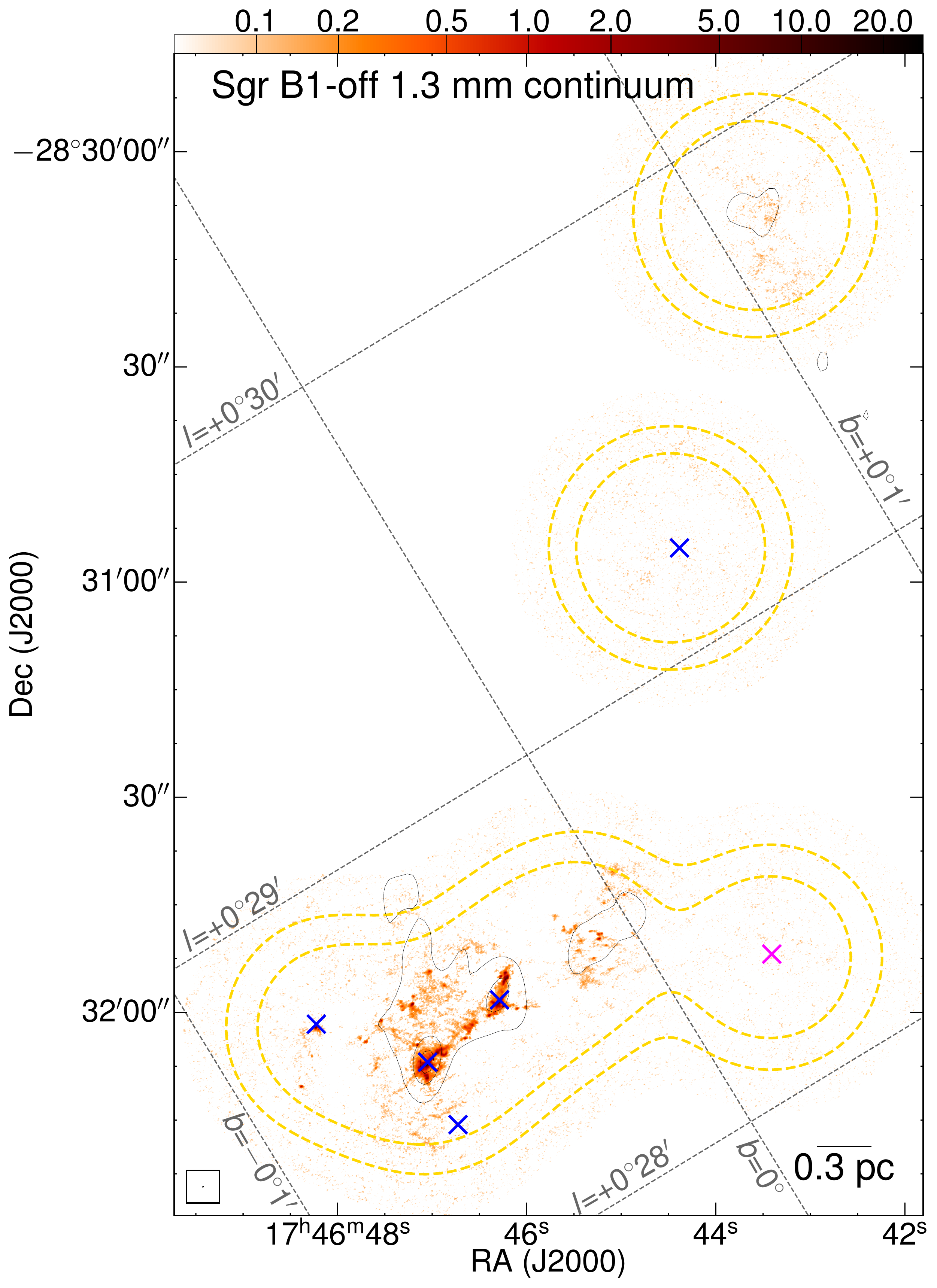} \\
\includegraphics[width=0.53\textwidth]{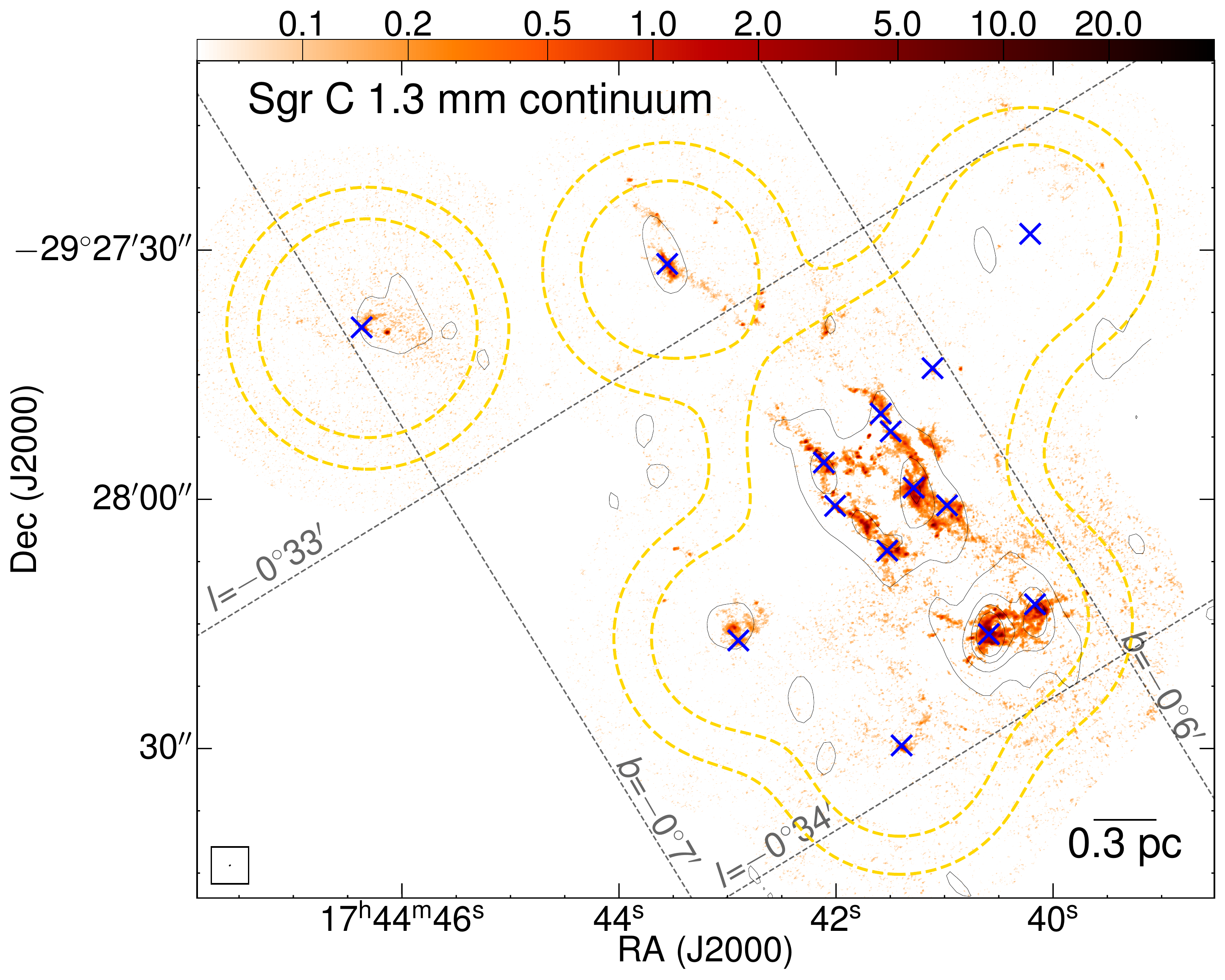}
\end{tabular}
\end{tabular}
\caption{ALMA 1.3~mm continuum emission of the four clouds is shown as color scale displayed on a logarithmic scale in units of \mjypbm{}. The inner and outer yellow dashed loops show the ALMA primary-beam responses at 50\% and 30\%, respectively. Black contours show the SMA 1.3~mm continuum emission with an angular resolution of 5\arcsec{}$\times$3\arcsec{} \citep{lu2019a}, starting from 5$\sigma$ in steps of 20$\sigma$ where $1\sigma$=3~\mjypbm{}. Crosses show the positions of \water{} masers, among which the magenta ones are known AGB stars \citep{lu2019a}. Diagonal dashed lines denote Galactic coordinates.}
\label{fig:cont}
\end{figure*}

Our ALMA observations targeted selected regions in the four clouds (\autoref{fig:cont}). All the regions either have gas masses of $\gtrsim$10$^3$~\msol{} within a radius of $\sim$0.5~pc or are associated with \water{} masers \citep{lu2019a}, and therefore are potential high-mass cluster forming sites. 

The data were taken in the C43-5 and C43-3 configurations (project code: 2016.1.00243.S), and were calibrated separately and then imaged together. The correlators were set to cover frequencies between 217--221~GHz and 231--235~GHz, with a uniform spectral resolution of 1.129~MHz (1.5~\kms{}). The calibration was done using the standard pipeline implemented in CASA 4.7.2. The imaging was done using CASA 5.4.0. Continuum emission was reconstructed from line-free channels, with a central frequency at 226~GHz. We used the \textit{tclean} task in CASA, with Briggs weighting and a robust parameter of 0.5, and a multiscale algorithm with scales of [0, 5, 15, 50, 150] and a pixel size of 0\farcs{04}. We further performed two iterations of phase-only self-calibrations for Sgr~B1-off, and two iterations of phase-only plus one iteration of phase \& amplitude self-calibrations for Sgr~C, where bright continuum sources exist, to improve imaging dynamic range. The time interval used for solving self-calibration solutions is the shortest integration duration (2.048 s).

The resulting synthesized beam is 0\farcs{25}$\times$0\farcs{17} (equivalent to 2000~AU$\times$1400~AU). Due to a lack of short baselines, the data are not sensitive to structures larger than 7\arcsec{} ($\sim$0.3~pc). The image rms measured in emission-free regions without primary-beam corrections is 40~\mujypbm{}, except around the brightest peak in Sgr~C where it is $\sim$60~\mujypbm{} after self-calibration.

%%%%%%%%%%%%%%%%%%%%%%%%%
\section{RESULTS}\label{sec:results}

\subsection{Identification of Cores}\label{subsec:results_fragmentation}
We studied gas fragmentation using the continuum, which is mostly contributed by thermal dust emission (see \citealt{lu2019a}, in which we concluded that the continuum at this frequency is dominated by cold dust emission). As shown in \autoref{fig:cont}, the four clouds exhibit distinctly different fragmentation levels. The 20~\kms{} cloud and Sgr~C have the most fragmented substructures. Sgr~B1-off shows moderate fragmentation in its southern part. The 50~\kms{} cloud shows little fragmentation. 

We used the dendrogram algorithm \citep{rosolowsky2008dendro} implemented with \textit{astrodendro}\footnote{\url{www.dendrograms.org}} to search for substructures at $\sim$2000~AU scales in the continuum images, and defined the identified leaves (the base element in the hierarchy of the dendrogram that has no further substructure) as cores. The 2000 AU-scale cores are smaller than those 0.2 pc-scale cores defined in our previous works \citep{lu2019a}. The algorithm was run on the images without primary-beam corrections that have uniform noise levels, allowing us to apply a single set of criteria to the whole image. Fluxes of the identified cores were taken from primary-beam corrected images. We set the minimum flux density to 4$\sigma$, the minimum significance for structures to 1$\sigma$, and the minimum area to the size of the synthesized beam. We dropped all leaves lying outside of the 30\% primary-beam response, where the sensitivity deteriorates significantly and the identified cores are biased to the brightest ones.

About 800 cores were identified in the 20~\kms{} cloud, Sgr~C, and the southern part of Sgr~B1-off. No cores were found in the 50~\kms{} cloud or the northern part of Sgr~B1-off, which is likely a result of strong turbulence (FWHM$\sim$8--12~\kms{}) at $\gtrsim$0.1~pc scales that hinders the formation of bound fragments \citep{lu2019a}. We thus excluded the two regions from the following discussions.

\begin{figure*}[!t]
\centering
\begin{tabular}{@{}p{0.45\textwidth}@{}p{0.55\textwidth}@{}}
\begin{tabular}[c]{@{}c@{}}
\includegraphics[width=0.45\textwidth]{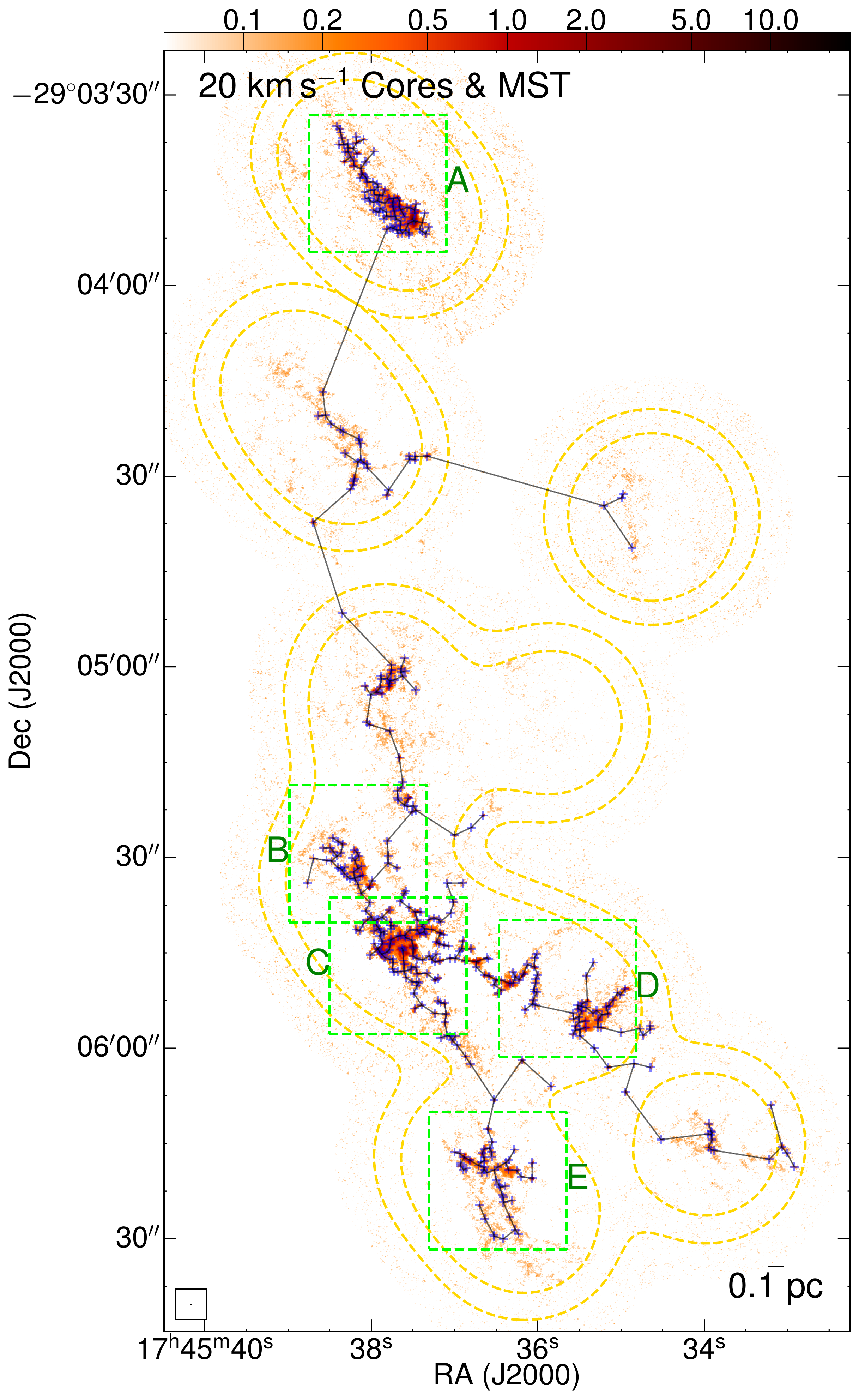}
\end{tabular}
&
\begin{tabular}[c]{@{}c@{}}
\includegraphics[width=0.41\textwidth]{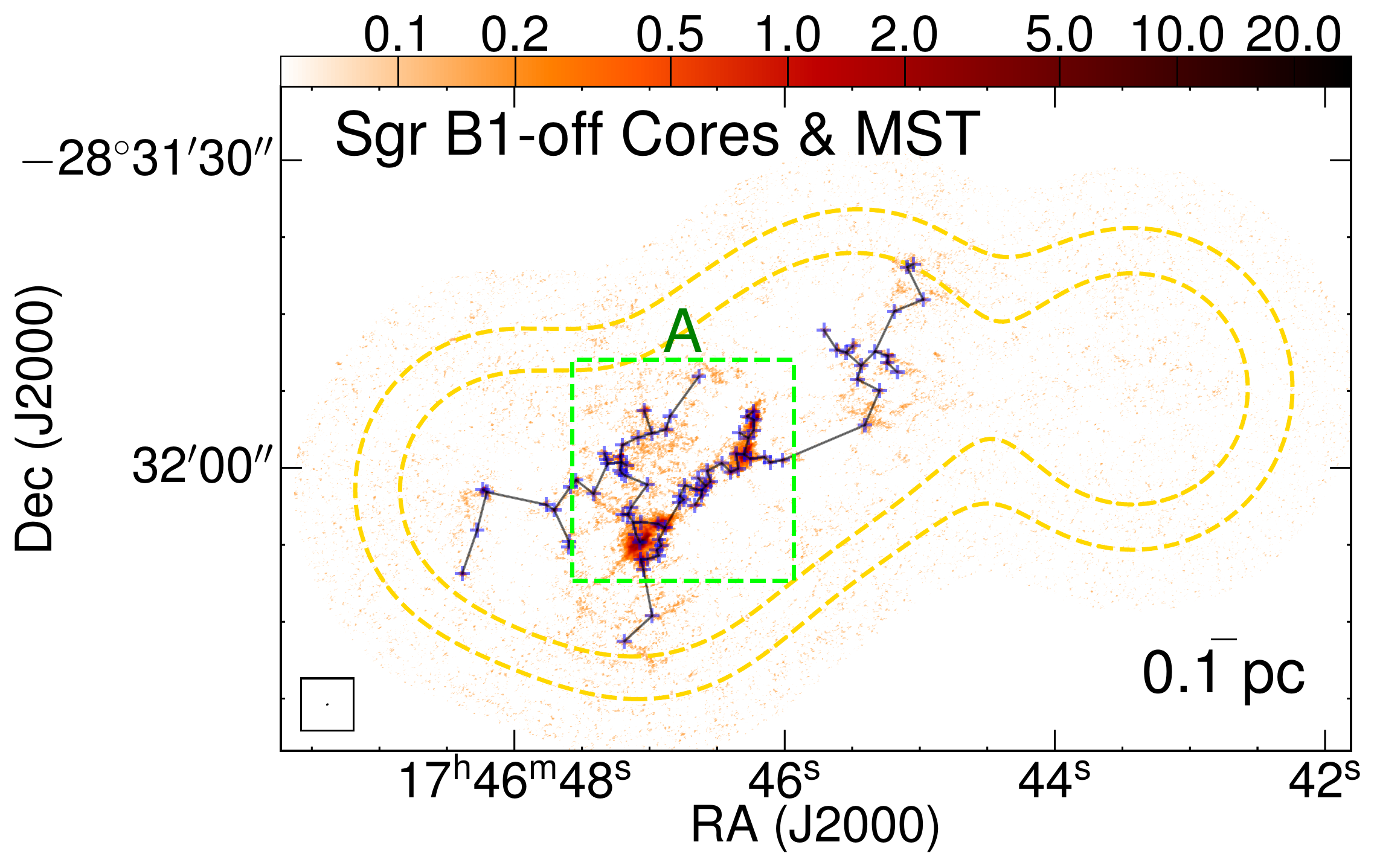} \\
\includegraphics[width=0.53\textwidth]{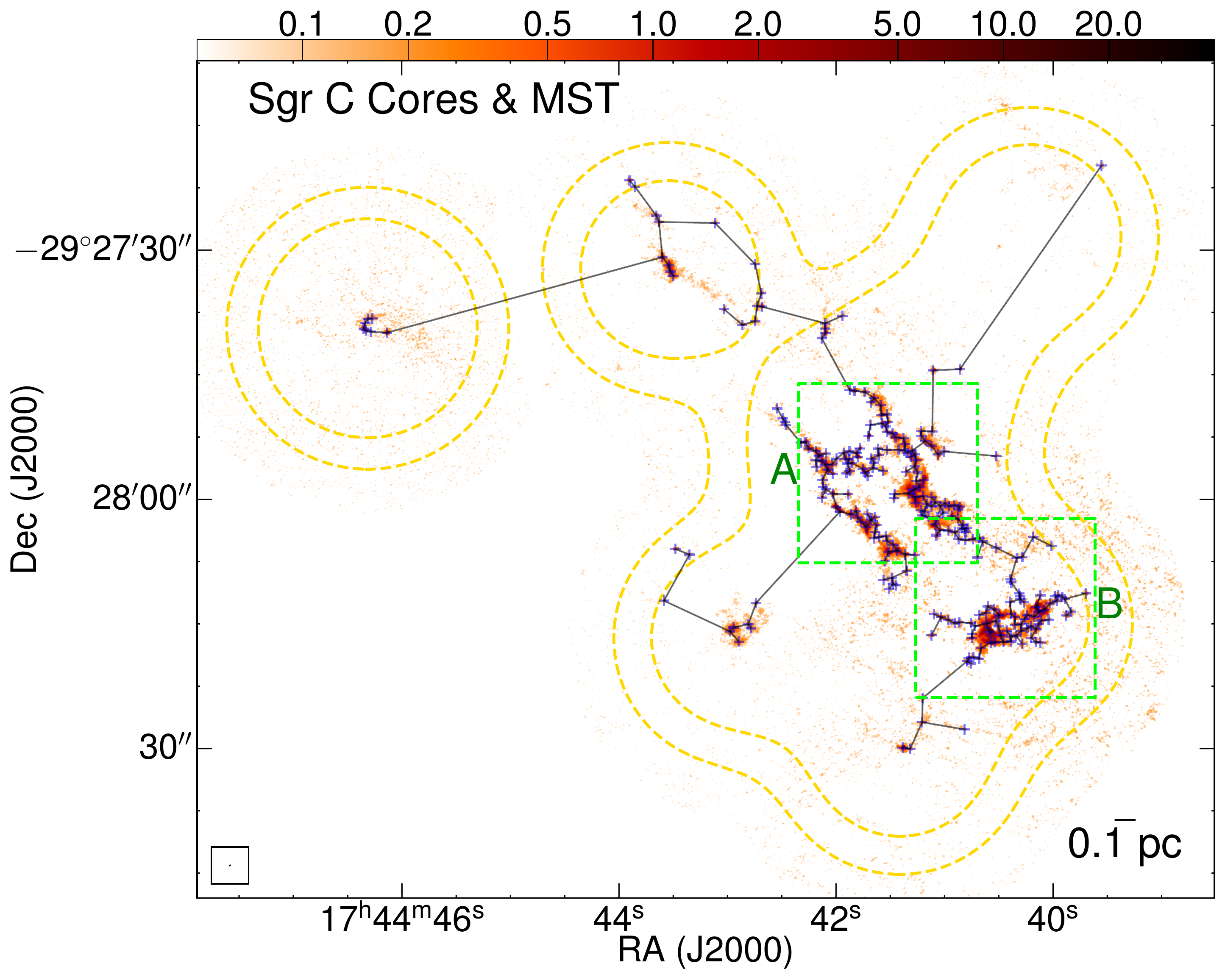}
\end{tabular}
\end{tabular}
\caption{Blue crosses show peak positions of the cores identified by the dendrogram (\autoref{subsec:results_fragmentation}), and black segments show the MSTs (\autoref{subsec:disc_jeans}). Color scale is the same as in \autoref{fig:cont}. Zoomed-in views of clustered cores in the green boxes are in \autoref{sec:appd_zoom}.}
\label{fig:dendro}
\end{figure*}

The identified cores are marked by crosses in \autoref{fig:dendro}, and zoomed-in views of clustered cores can be found in \autoref{sec:appd_zoom}. The full core catalog is available in \autoref{sec:appd_catalog} as a machine-readable table. In \autoref{sec:appd_dendro}, we explore the impact of varying the dendrogram parameters upon the following analyses (e.g., changing the minimum significance to 2$\sigma$) and find that it does not affect our conclusions.

\subsection{Physical Properties of the Cores}\label{subsec:results_properties}
Assuming optically thin dust emission, the core masses were derived as
\begin{equation}\label{equ:coremass}
M_\text{c}=R\frac{S_\nu d^2}{B_\nu(T_\text{dust})\kappa_\nu},
\end{equation}
where $R$ is the gas-to-dust mass ratio, $S_\nu$ is the dust emission flux, $d$ is the distance, $B_\nu(T_\text{dust})$ is the Planck function at the dust temperature $T_\text{dust}$, and $\kappa_\nu$ is the dust opacity. We assumed $R$=100 and $\kappa_\nu$=0.899~cm$^2$\,g$^{-1}$ \citep[MRN model with thin ice mantles, after 10$^5$ years of coagulation at 10$^6$~\cc{};][]{ossenkopf1994}. 

$T_\text{dust}$ at 2000 AU scales is unclear. The dust temperature at $\gtrsim$0.5~pc in these clouds is measured to be 20~K based on \textit{Herschel} observations \citep{kauffmann2017a}. At $\sim$0.1~pc scales, the gas temperature is found to be 50--200~K and higher \citep{mills2013,lu2017,walker2018}. However, it is unclear whether the dust and gas at 0.1~pc (and smaller) scales are in thermodynamic equilibrium. For simplicity, we assumed a constant $T_\text{dust}$ of 20~K as a fiducial case. In \autoref{sec:appd_tdust}, we will discuss the effect of different dust temperatures and show that it is significant to core masses. For example, if a higher dust temperature of 50~K was adopted, the core masses would decrease by a factor of 3. Future high angular resolution multi-wavelength observations will be critical for resolving the dust temperature ambiguity.

The effective core radius $r_\text{c}$ is derived as $(A/\pi)^{1/2}$, where $A$ is the area of the core reported by \textit{astrodendro}. The molecular hydrogen volume density $n(\text{H}_2)$ is then $M_\text{c}/(4\pi r_\text{c}^3/3)/(2.8m_\text{H})$.

We reported statistics of the fluxes, masses, radii, and densities of the cores in \autoref{tab:cores}. Note that at small scales of 2000~AU, the missing flux issue of ALMA as an interferometer unlikely affects the measurement of fluxes. With a continuum emission rms of 40~\mujypbm{}, the 5$\sigma$ mass sensitivity is 0.3~\msol{} per beam given a $T_\text{dust}$ of 20~K.

About 30 cores are spatially associated with \water{} masers (\autoref{fig:cont}) or UC \hii{} regions \citep{lu2019a}, therefore are likely protostellar, although the 3\arcsec{} resolution of observations in \citet{lu2019a} prevents us from assigning the star formation signatures to a particular core. The other cores are not associated with signatures of high-mass star formation found in previous observations \citep[e.g.,][]{kauffmann2017a,lu2019a,lu2019b}. However, we cannot rule out the possibility of low or intermediate-mass star formation associated with them. Their densities ($10^{6\text{--}8}$~\cc{}, \autoref{tab:cores}) are comparable to or greater than the critical density for star formation in the CMZ predicted by several studies \citep[$\sim$$10^7$~\cc{};][]{kruijssen2014,rathborne2014b,federrath2016b,kauffmann2017b,ginsburg2018a}. The free-fall time based on the densities is 3$\times$$10^3$--3$\times$$10^4$ years. The cores will likely end up with star formation, but with the current data we cannot determine whether the cores have collapsed and are mostly protostellar, or they have recently condensed out of the clouds in the last $\sim$$10^4$ years and are mostly prestellar.

\begin{deluxetable*}{cccccccccccccc}
\tabletypesize{\scriptsize}
\tablecaption{Observed core properties.\label{tab:cores}}
\tablewidth{0pt}
\tablehead{
\colhead{Cloud} & Number of & \multicolumn{3}{c}{Flux (mJy)} & \multicolumn{3}{c}{Mass (\msol)} & \multicolumn{3}{c}{Radius (AU)} & \multicolumn{3}{c}{Density ($10^7$~\cc{})} \\
 & cores & Range & Mean & Median & Range & Mean & Median & Range & Mean & Median & Range & Mean & Median
 }
\startdata
20~\kms{} & 471 & 0.20--131 & 2.9 & 0.9 & 0.31--198 & 4.4 & 1.4 & 1030--5970 & 1860 & 1620 & 0.28--23 & 1.5 & 0.9 \\
Sgr~B1-off          & 89    & 0.19--153 & 4.4 & 0.9 & 0.29--230  & 6.6 & 1.3 & 1030--8160 & 1880 & 1690 & 0.41--4.3 & 1.2 & 0.9 \\
Sgr~C          & 275  & 0.21--202 & 4.5 & 1.2 & 0.32--304 & 6.8 & 1.8 & 1030--6560 & 1800 & 1540 & 0.33--28 & 2.1 & 1.2
\enddata
\end{deluxetable*}

\begin{figure*}[!t]
\centering
\begin{tabular}{@{}cc@{}}
\includegraphics[width=0.45\textwidth]{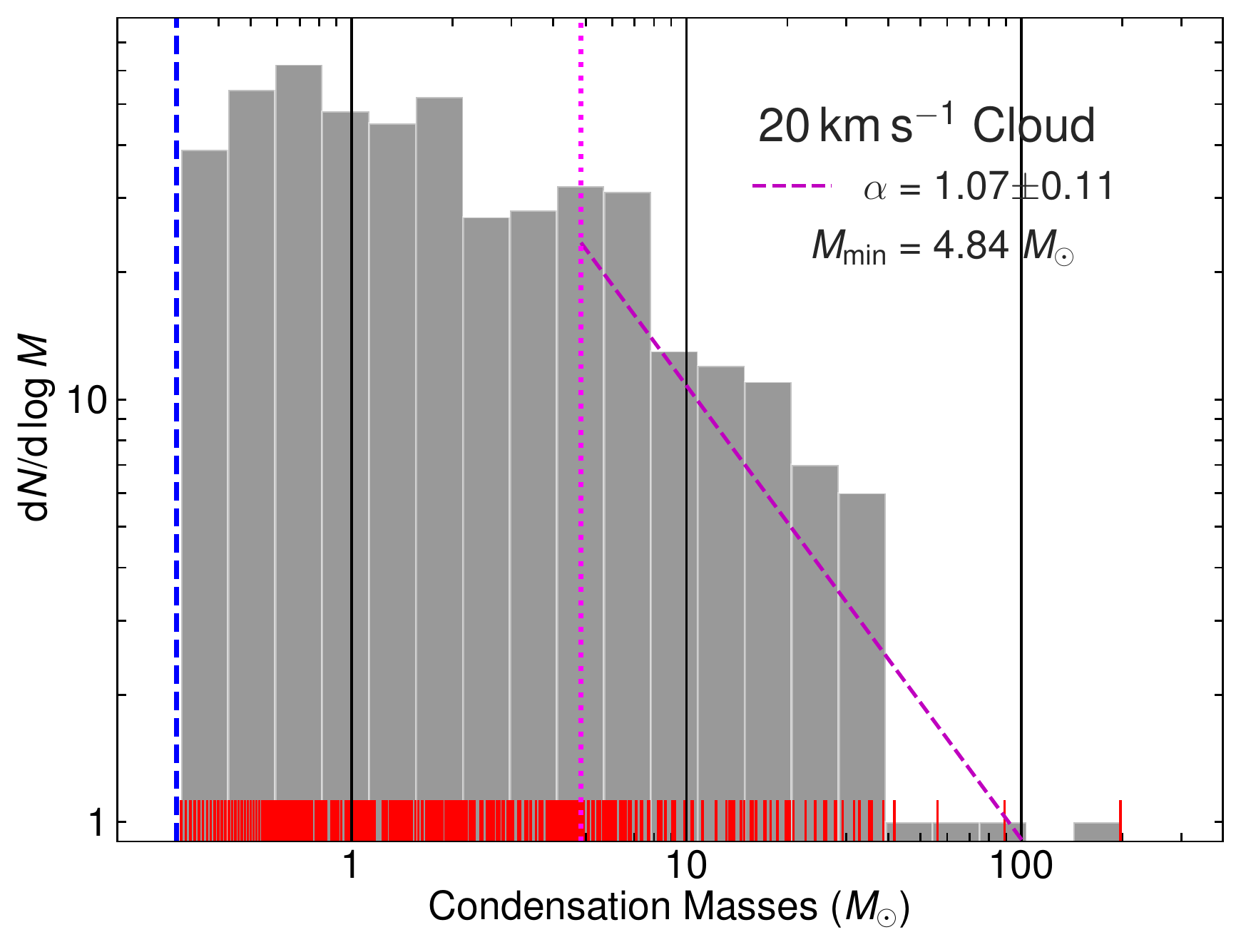} &
\includegraphics[width=0.45\textwidth]{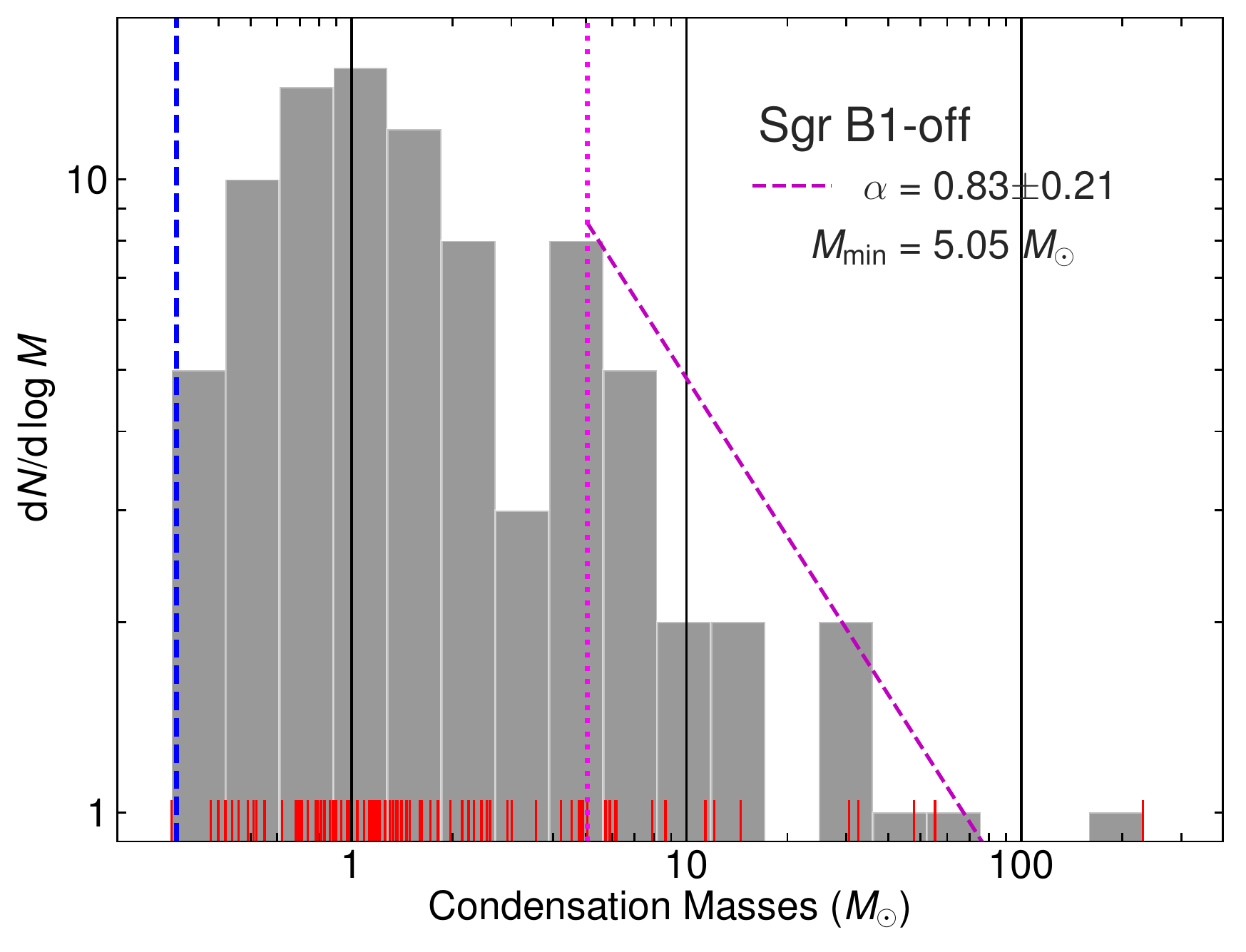} \\
\includegraphics[width=0.45\textwidth]{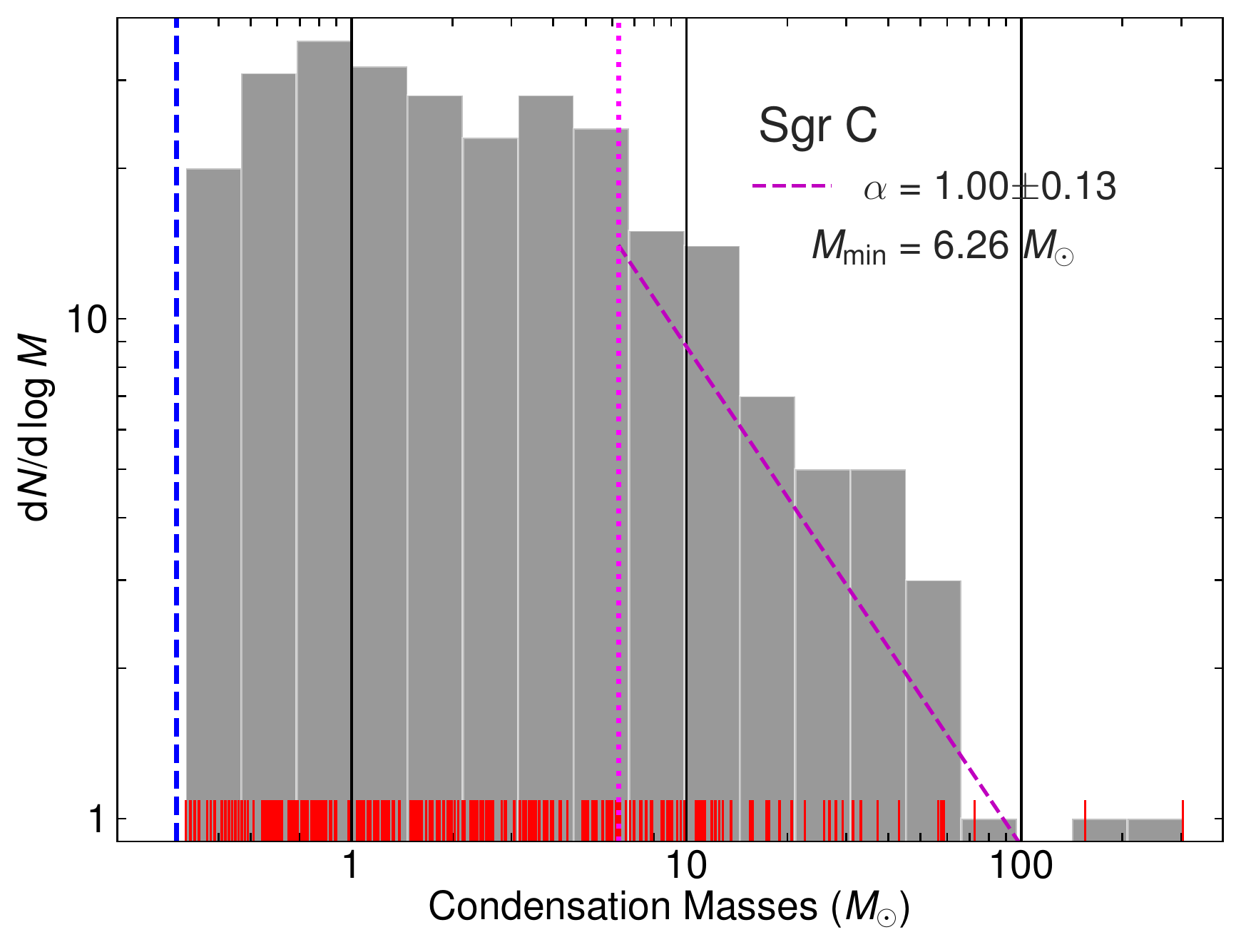} &
\includegraphics[width=0.45\textwidth]{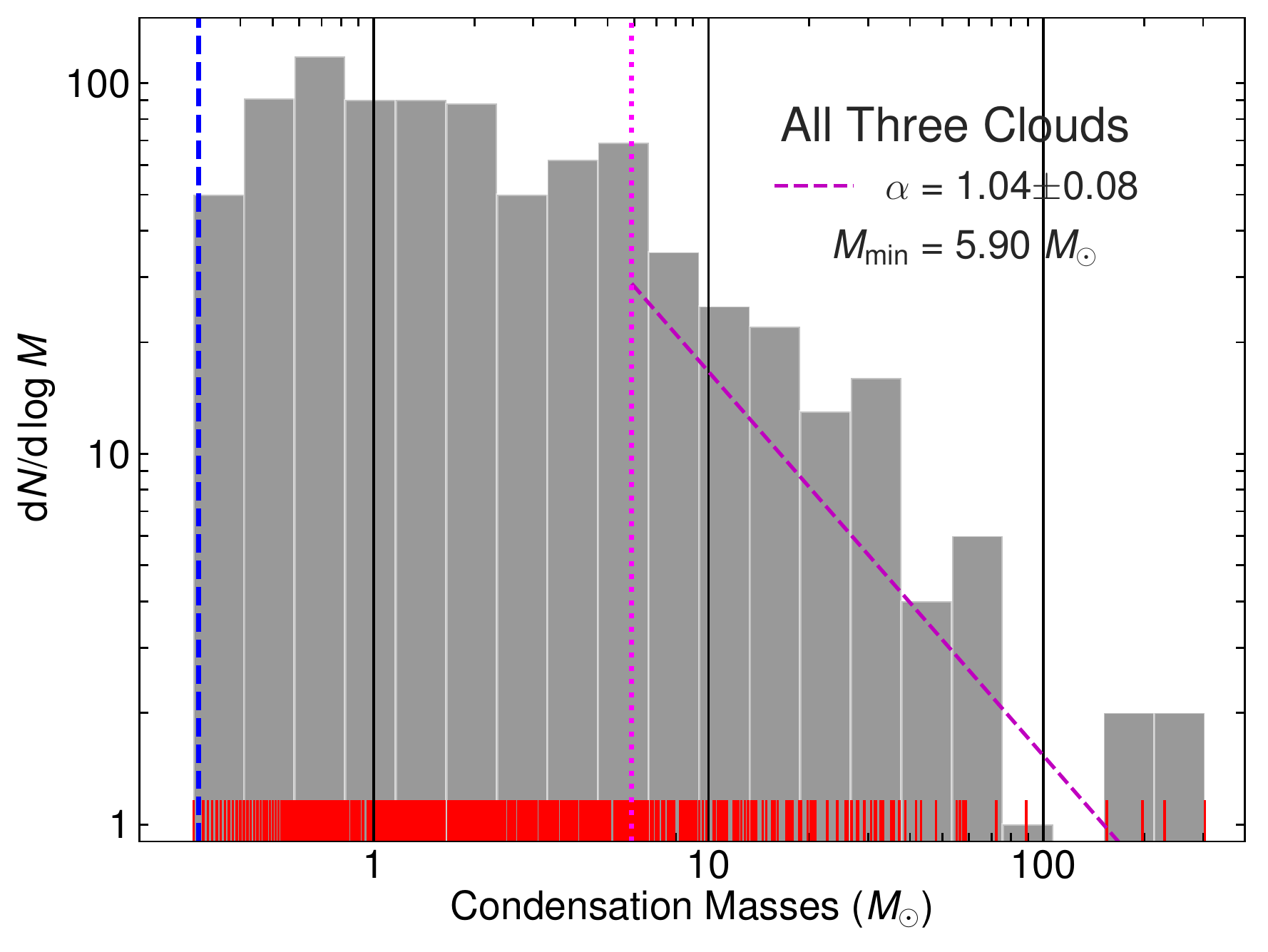}
\end{tabular}
\caption{CMFs of the three individual clouds and the three clouds together. The red sticks are the actual data points. The vertical blue line marks the 5$\sigma$ mass sensitivity (0.3~\msol{}) provided a dust temperature of 20~K. The magenta dashed line is not a fit to the histograms, but represents the result of MLE plus an arbitrary normalization factor. The vertical magenta line denotes the lower bound given by the MLE method.}
\label{fig:cmf}
\end{figure*}

Using the core masses, we constructed core mass functions (CMFs) of the three individual clouds as well as all three clouds combined, as presented in \autoref{fig:cmf}. We fit the high-mass end of the CMFs with a power law function:
\begin{equation}
\frac{\text{d}\,N}{\text{d}\,\log\,M}\propto M^{-\alpha},
\end{equation}
using the maximum likelihood estimation (MLE) method in \citet{clauset2009} implemented with the \textit{plfit} package\footnote{\url{https://github.com/keflavich/plfit}}. The method simultaneously fits the lower bound and the power-law index. The results are labeled in \autoref{fig:cmf}.

The fitting to the CMF of Sgr~B1-off is less robust as there are less cores. The other three fittings reach similar values of $\alpha$ in the range of 1.00--1.07. The lower bound of all fittings is found to be around 5~\msol{}, which can be explained by the confusion limit in clustered environments provided that the core masses are drawn from a power law distribution (\autoref{sec:appd_lowerbound}).

The relation between the CMF and the initial mass function (IMF) is still under debate. The high-mass end ($\geq$0.5~\msol{}) of the IMF has been fit using a power law with an index of $\alpha$=$1.3$ \citep{kroupa2001} and is commonly assumed to be universal (however, see e.g.\ \citealt{hopkins2018}). Our result suggests a slightly shallower power-law index for the CMFs, which is similar to recent findings toward clouds in the Galactic disk \citep[e.g.,][]{motte2018,liu2018,sanhueza2019}. However, there are several significant uncertainties. As shown in Appendices \ref{sec:appd_dendro} \& \ref{sec:appd_tdust}, depending on the adopted dendrogram parameters or dust temperatures, the power-law index could vary significantly. We stress that great caution must be taken to interpret the power law in the CMFs, and our result and similar studies toward Galactic disk clouds in the literature, although being consistent with each other, suffer from the same uncertainties.

%%%%%%%%%%%%%%%%%%%%%%%%%
\section{DISCUSSION}\label{sec:disc}
\subsection{Thermal Jeans Fragmentation}\label{subsec:disc_jeans}

\begin{figure*}[!t]
\centering
\begin{tabular}{@{}cc@{}}
\includegraphics[width=0.45\textwidth]{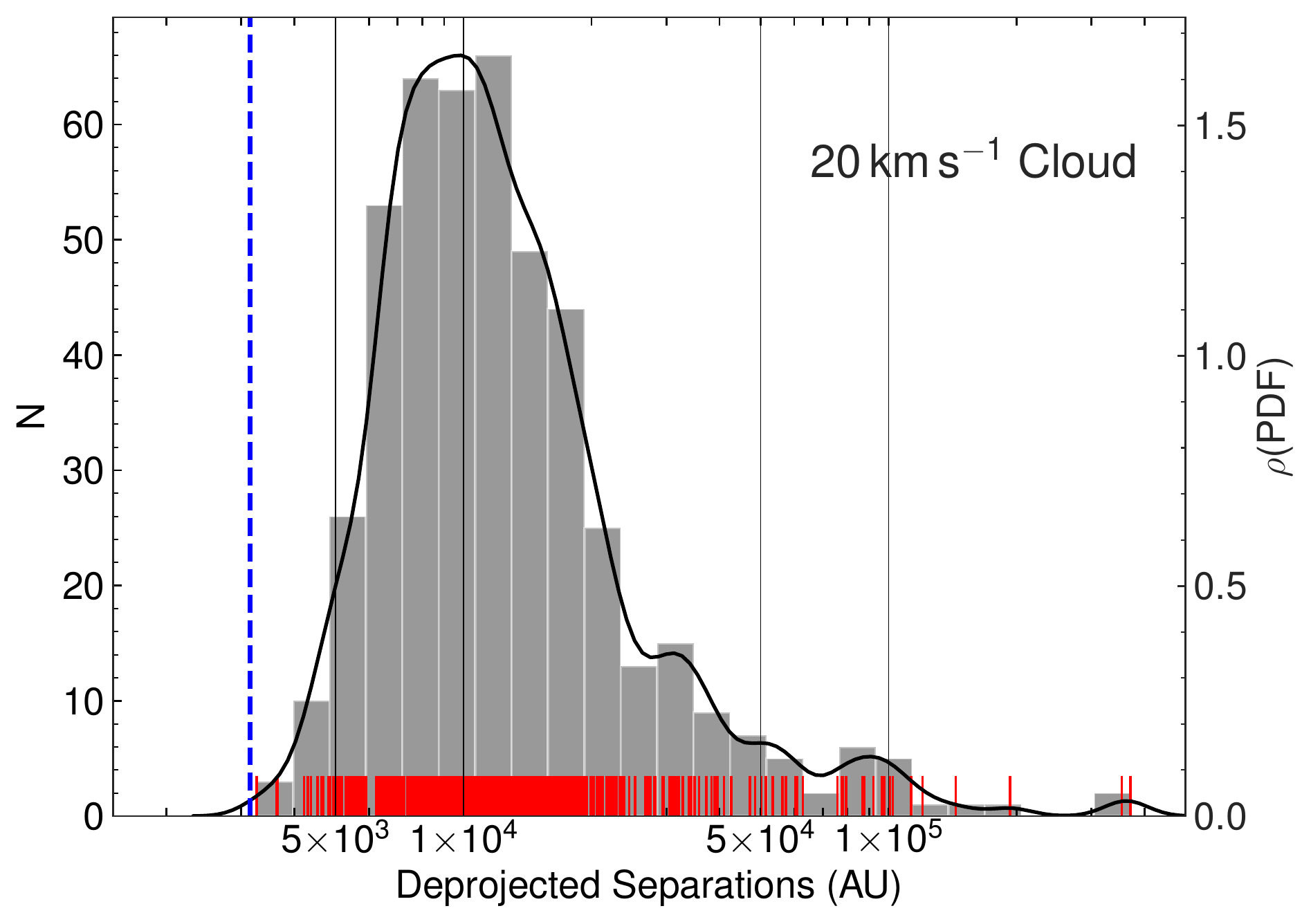} &
\includegraphics[width=0.45\textwidth]{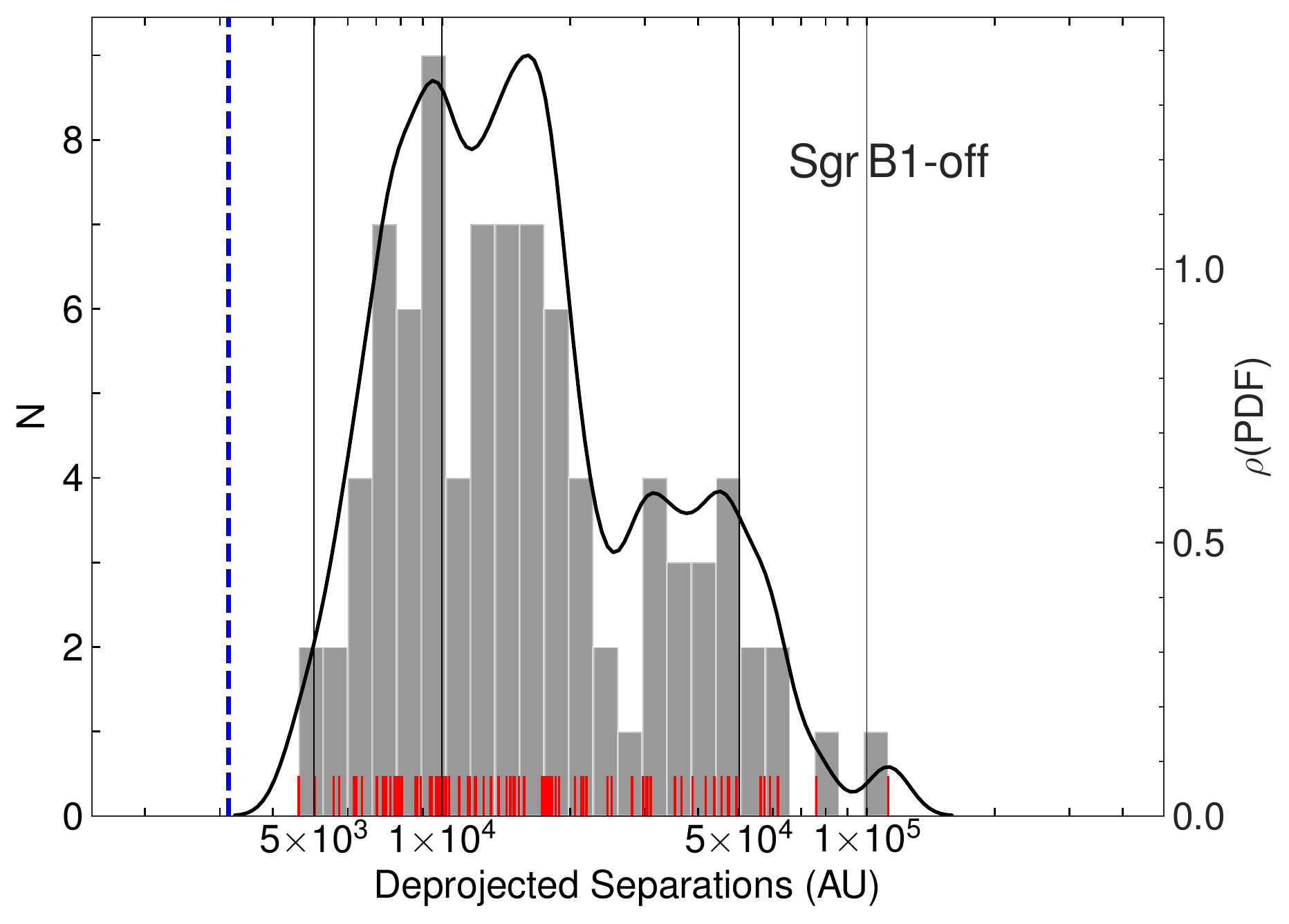} \\
\includegraphics[width=0.45\textwidth]{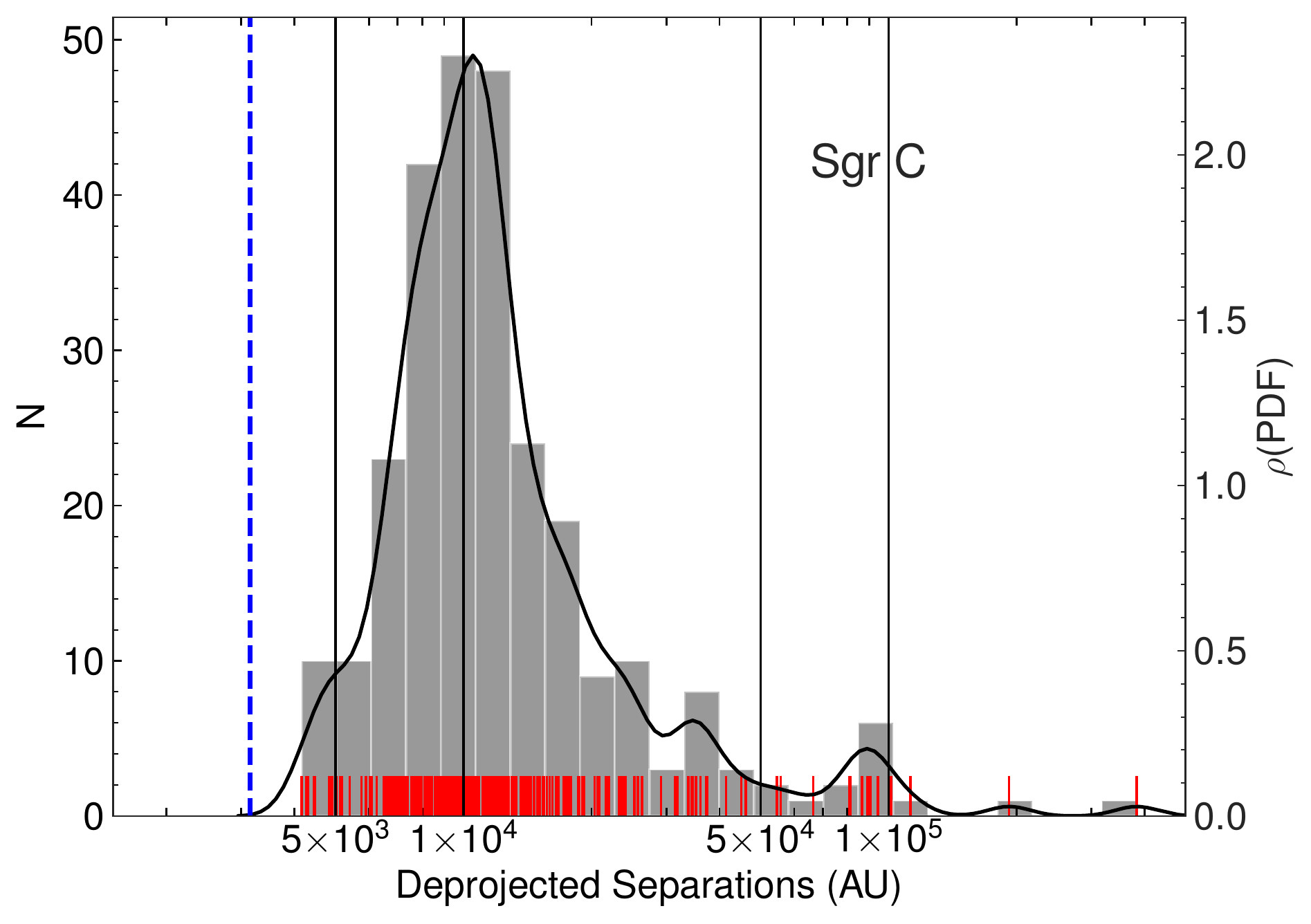} &
\includegraphics[width=0.45\textwidth]{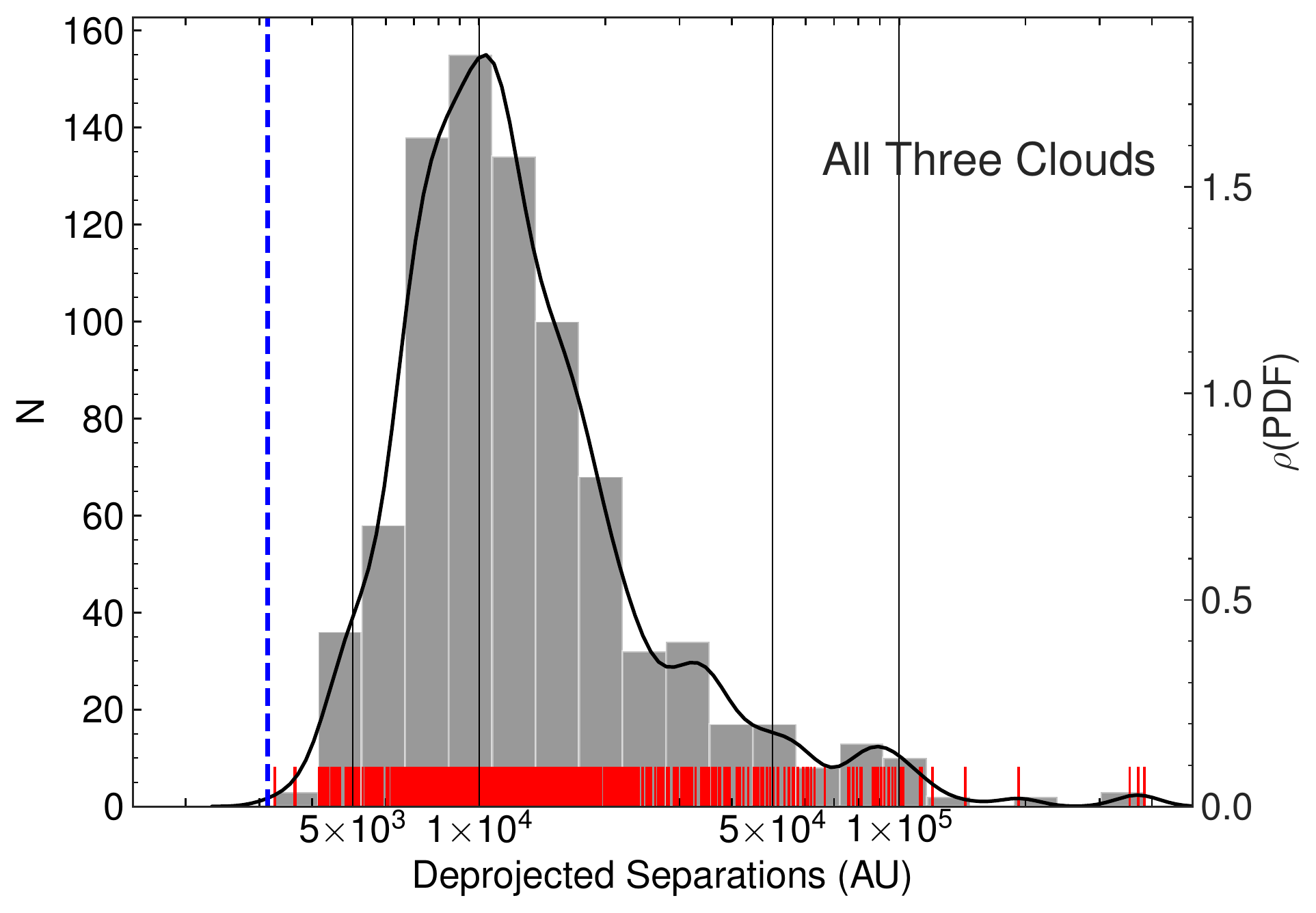}
\end{tabular}
\caption{Distributions of deprojected spatial separations between the cores. In each panel, the red sticks, the gray bars, and the black curves are the actual data points, the histogram, and the kernel density estimate (KDE) of the same distribution. Labels on the left-side vertical axis mark numbers of data points in the histogram bins, while those on the right-side mark values of the probability density function for the KDE, with the probability of a certain range of separations being the area under the curve in that range. The vertical blue dashed line marks the spatial resolution (0\farcs{25}$\sim$2000 AU) divided by the projection factor 2/$\pi$.}
\label{fig:separations}
\end{figure*}

Previous studies have suggested that turbulence with line widths of 5--10~\kms{} dominates the gas dynamics from $\gtrsim$1~pc to $\gtrsim$0.1~pc scale in the CMZ, potentially leading to the emergence of massive clouds on the one hand and inhibiting gas collapse in the clouds on the other hand \citep{federrath2016b,henshaw2016,kruijssen2019a}. At 0.1~pc scales, recent high spatial resolution observations find smaller line widths of $\lesssim$1~\kms{} \citep{kauffmann2017a,barnes2019}. With such narrow line widths, it is unclear whether the gas dynamics are still dominated by turbulence.

Here we investigate the gas fragmentation at sub-0.1~pc scales and compare with thermal Jeans fragmentation. If the observation is consistent with thermal Jeans fragmentation, it may suggest that the strong turbulence on larger scales has decayed to allow for active star formation on smaller scales.

To study spatial scales in the fragmentation, we apply the minimum spanning tree (MST) algorithm to the cores. This algorithm calculates the sum of edge lengths that connect nodes without any loop in a graph, and finds the minimum value as well as the corresponding edge collection, defined as the MST. In our case, the nodes are the cores, and the edge lengths are the projected spatial separations between the cores.

The MST algorithm is implemented with a modified version of the FragMent package \citep{clarke2019}. The MSTs of the clouds are shown as black segments in \autoref{fig:dendro}. To correct for the projection effect from the 3D space to the 2D sky, we divide the edge lengths by a factor of 2/$\pi$ to get the deprojected separations \citep{sanhueza2019}. Note that owing to incomplete spatial sampling (e.g., isolated pointings), longer separations are generally not meaningful. Here we focus on short separations ($\lesssim$1$\times$$10^5$~AU, half of the FWHM primary beam size) which are not affected. 

Distributions of the deprojected separations are plotted in \autoref{fig:separations}. In the three clouds, the most frequent separation is $\sim$$10^4$~AU. If the dendrogram parameters are different, the most frequent separation can vary between (0.8--1.5)$\times$$10^4$~AU (\autoref{sec:appd_dendro}).

We then compare the spatial separations with Jeans fragmentation. When a piece of homogeneous gas undergoes fragmentation with thermal pressure, the characteristic separation between the fragments is described by the Jeans length:
\begin{equation}\label{equ:lj}
\lambda_\text{J} = c_\text{s}\left(\frac{\pi}{G\rho}\right)^{0.5},
\end{equation}
where $c_\text{s}$ is the isothermal sound speed, and $\rho$ is the density of the gas that can be derived as $n(\text{H}_2)$$\times$$2.37m_\text{H}$.

The parental gas from which these 2000 AU-scale cores arise through hierarchical fragmentation is the 0.2 pc-scale cores, which have been studied in \citet{lu2017,lu2019a}. As such, we adopt the characteristic gas temperature 50--200~K and gas density $10^6$~\cc{} from those works. Note that the temperature is that of the gas at 0.2~pc scales, which is different from that of the dust at 2000~AU scales adopted for \autoref{equ:coremass}. The derived Jeans length is (1.0--1.9)$\times$$10^4$~AU. Therefore, the observed core separations are consistent with thermal Jeans length.

Jeans fragmentation also predicts a characteristic fragment mass defined as the Jeans mass:
\begin{equation}\label{equ:mj}
M_\text{J} = \frac{4\pi\rho}{3}\left(\frac{\lambda_\text{J}}{2}\right)^3=c_\text{s}^3\frac{\pi^{5/2}}{6\sqrt{G^3\rho}}.
\end{equation}
We compare the core masses and the Jeans masses, although this comparison is less robust than that between the separations considering the uncertainties in the estimate of core masses. As shown in \autoref{tab:cores}, the characteristic mass based on mean or median values is 1--7~\msol{}. This is generally consistent with the Jeans mass of 3--25~\msol{}. There are cores with larger masses (up to $\gtrsim$100~\msol{} with the current assumptions). These cores may form by turbulent-supported fragmentation \citep{hennebelle2008,zhang2009,wang2014} or have accumulated their masses through further gas accretion.

Recent high angular resolution observations toward massive clouds in the Galactic disk have revealed thermal Jeans fragmentation at sub-0.1 pc scales \citep{palau2018,liu2019,sanhueza2019}. The same scenario is likely controlling the fragmentation at sub-0.1~pc scales in the highly turbulent CMZ clouds, in contrast to the situation at larger scales in the CMZ ($\gtrsim$0.1~pc) where turbulence dominates the gas dynamics.

\subsection{Formation of Star Clusters}\label{subsec:disc_clusters}
The CMFs in \autoref{fig:cmf} show that each of the three clouds contains 5--20 cores above $\gtrsim$20~\msol{}. Assuming a star formation efficiency of at least 50\% and no further fragmentation, all of these cores will give rise to high-mass stars. Considering also the fact that these clouds are only marginally bound \citep{kauffmann2017b}, we expect that these clouds will at most form small OB associations with $\leq$20 high-mass stars. On the other hand, Arches and Quintuplet each contains about 100 O-type stars in a small radius of $\sim$1~pc \citep{luj2018}. Therefore, none of the clouds in our observations are able to form Arches/Quintuplet-like clusters with the current population of cores \citep[see also][]{walker2016}.

In \autoref{sec:appd_tdust} we discuss the impact of dust temperatures to the core masses. If a higher dust temperature is assumed, the masses will be smaller and there will be even fewer cores that are able to form high-mass stars.

There is a possibility for these clouds to form Arches/Quintuplet-like clusters if the less massive cores continue accreting gas and grow heavier to form high-mass stars. The clouds have sufficient gas reservoir ($\gtrsim$$10^5$~\msol{}) to feed into the cores and give birth to a cluster of $\gtrsim$$10^4$~\msol{} assuming an overall efficiency of 2--10\% \citep{kruijssen2019b,chevance2020}. If a 2~\msol{} (median value in \autoref{tab:cores}) core accumulates gas at an average accretion rate of $\sim$2$\times$10$^{-3}$~\msolpyr{} for a free-fall timescale of $\sim$10$^4$ years, it will grow to $\gtrsim$20~\msol{} and may form a high-mass star. Such a high accretion rate has been observed toward prestellar cores in Galactic disk clouds \citep[e.g.,][]{contreras2018}. Virial analysis of several CMZ clouds also suggests evidence of global gravitational collapse \citep{barnes2019}. Future observations that aim to investigate gas accretion around the cores (e.g., using infall signatures seen in optically thick HCN/\hcop{} lines) will help examine this possibility.

%%%%%%%%%%%%%%%%%%%%%%%%%
\section{CONCLUSIONS}\label{sec:conclusions}
High angular resolution ALMA observations toward a sample of four massive clouds in the CMZ reveal hundreds of 2000 AU-scale cores. A power-law fit to the high-mass end of the CMFs suggests a slightly top-heavy shape ($\alpha$=0.83--1.07) as compared to the canonical IMF, which is similar to results toward Galactic disk clouds, but the fitting is highly susceptible to several uncertainties, e.g., the dust temperatures. Characteristic spatial separations and masses of the cores are consistent with thermal Jeans fragmentation. These results may imply similar star formation processes at sub-0.1~pc scales in the highly turbulent CMZ and elsewhere in the Galaxy, modulated by thermal Jeans fragmentation and leading to similar CMF shapes. Despite the fact that these are some of the most massive clouds and some of the only known sites of high-mass star formation in the CMZ, they are currently unable to form Arches/Quintuplet-like clusters, but may form such clusters by further gas accretion and core growth.

\acknowledgments
We thank the anonymous referee for constructive comments. XL thanks his family, Qinyu E and Xiaoe Lyu, for their support during the COVID-19 outbreak during which this manuscript was prepared, and Patricio Sanhueza, Benjamin Wu, and Hauyu Baobab Liu for helpful discussions. XL was supported by JSPS KAKENHI grant No.~JP18K13589. SNL thanks his family for their support during the COVID-19 outbreak. JMDK gratefully acknowledges funding from the Deutsche Forschungsgemeinschaft (DFG, German Research Foundation) through an Emmy Noether Research Group (grant number KR4801/1-1) and the DFG Sachbeihilfe (grant number KR4801/2-1), as well as from the European Research Council (ERC) under the European Union's Horizon 2020 research and innovation programme via the ERC Starting Grant MUSTANG (grant agreement number 714907). CB gratefully acknowledges support from the National Science Foundation under Award No.~1816715. This paper makes use of the following ALMA data: ADS/JAO.ALMA\#2016.1.00243.S. ALMA is a partnership of ESO (representing its member states), NSF (USA) and NINS (Japan), together with NRC (Canada), MOST and ASIAA (Taiwan), and KASI (Republic of Korea), in cooperation with the Republic of Chile. The Joint ALMA Observatory is operated by ESO, AUI/NRAO and NAOJ. Data analysis was in part carried out on the open-use data analysis computer system at the Astronomy Data Center (ADC) of NAOJ. This research has made use of NASA's Astrophysics Data System.

\vspace{5mm}

\software{CASA \citep{mcmullin2007}, astrodendro (\url{www.dendrograms.org}), FragMent \citep{clarke2019}, APLpy \citep{aplpy2012}, Astropy \citep{astropy2018}, searborn (\url{https://seaborn.pydata.org})}

\appendix
\section{Zoomed-in Views of Clustered Cores}\label{sec:appd_zoom}
The zoomed-in views of clustered and crowded subregions that are marked by green boxes in \autoref{fig:dendro} are displayed in \autoref{appd_fig:zooma}.

\begin{figure*}[!t]
\centering
\begin{tabular}{@{}p{0.45\textwidth}@{}p{0.45\textwidth}@{}}
\includegraphics[width=0.45\textwidth]{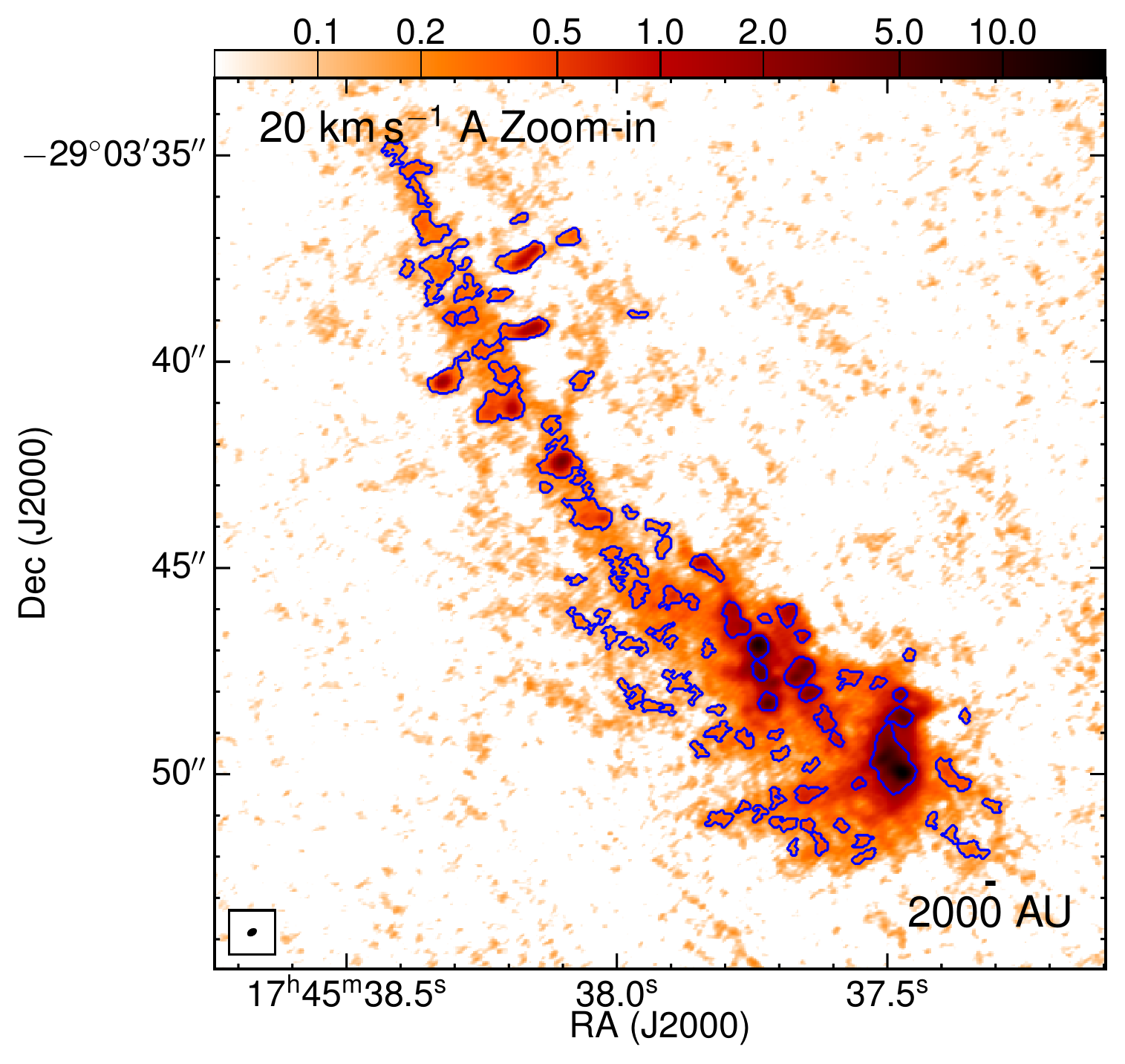} &
\includegraphics[width=0.45\textwidth]{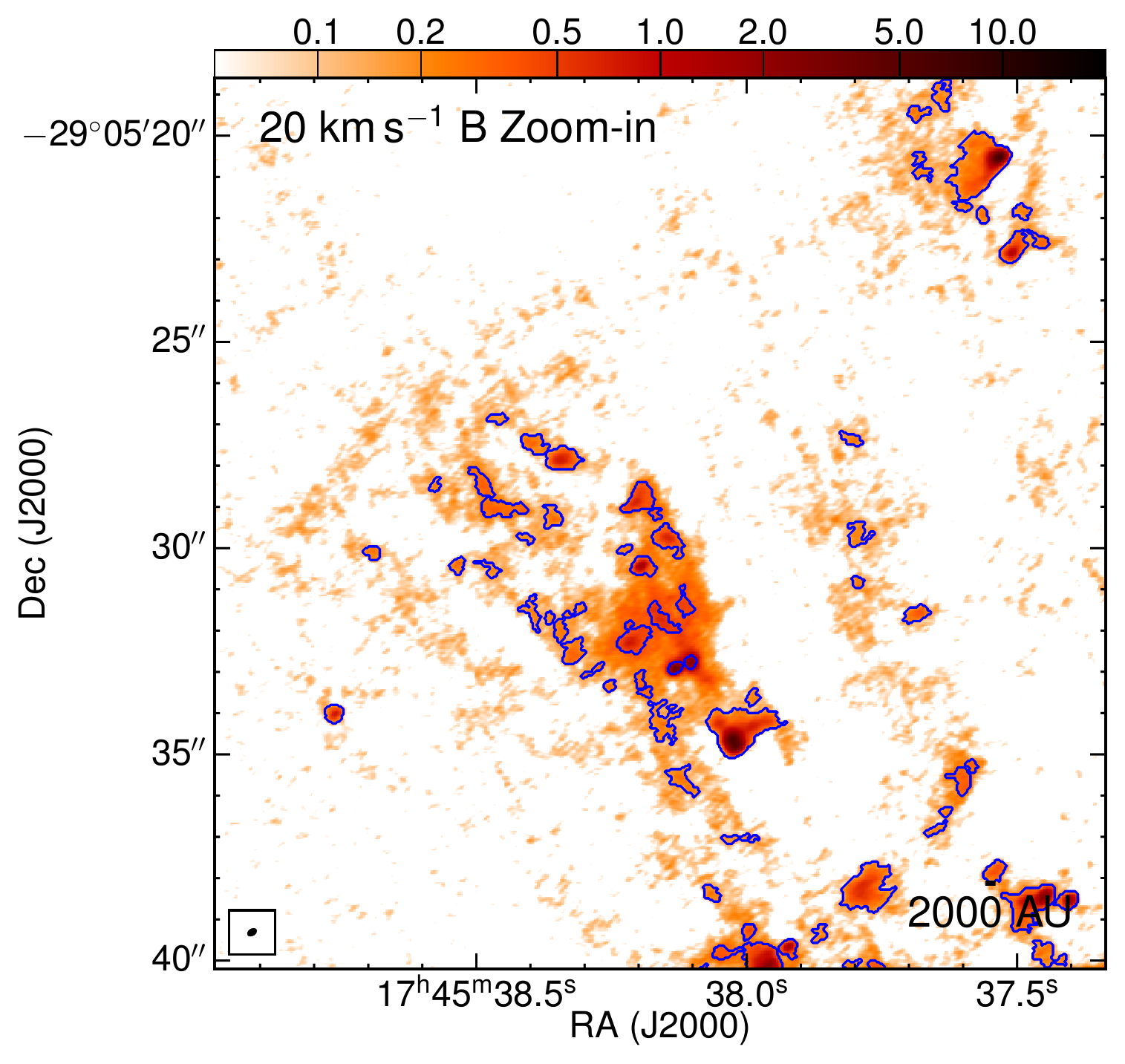} \\
\includegraphics[width=0.45\textwidth]{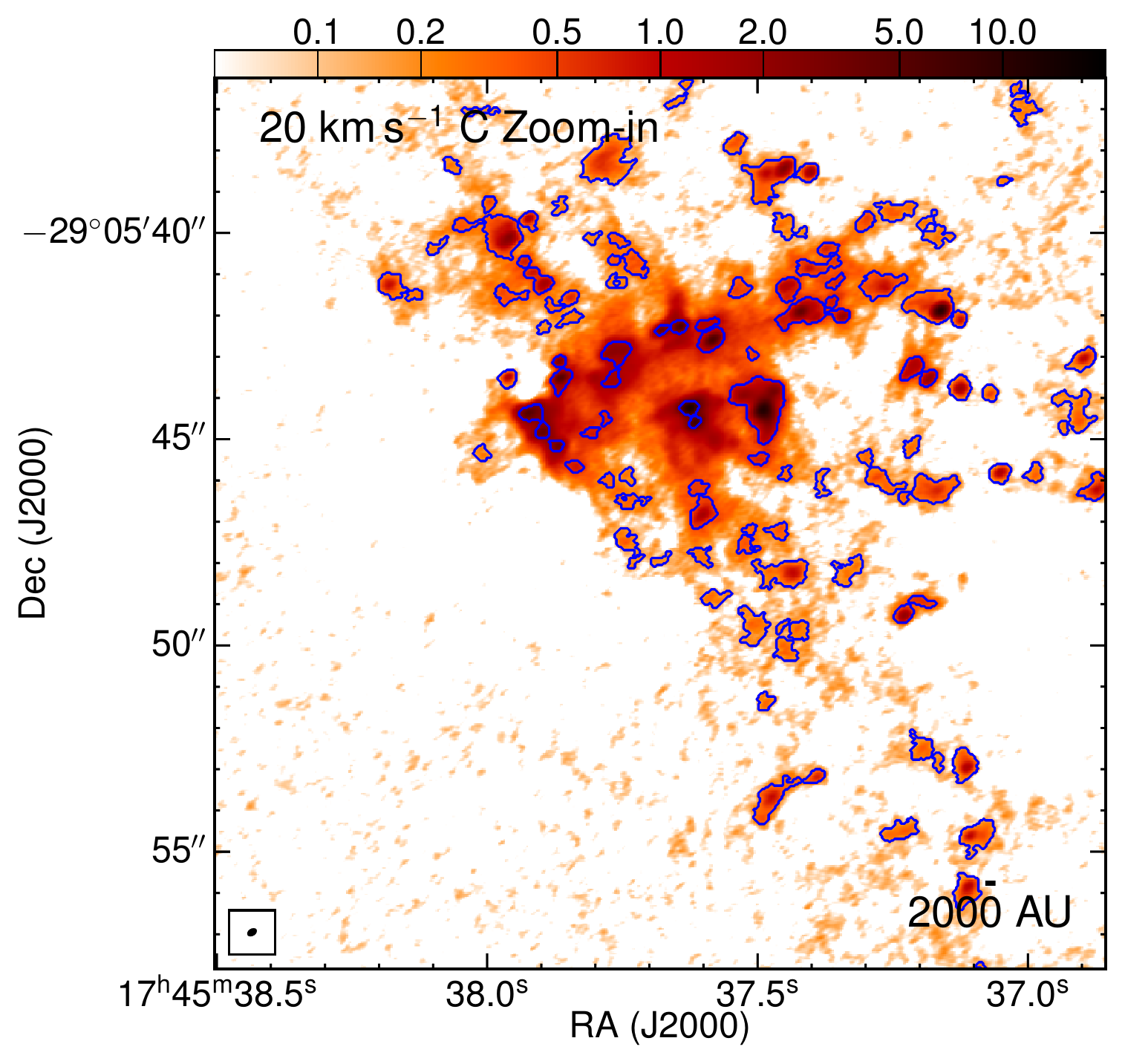} &
\includegraphics[width=0.45\textwidth]{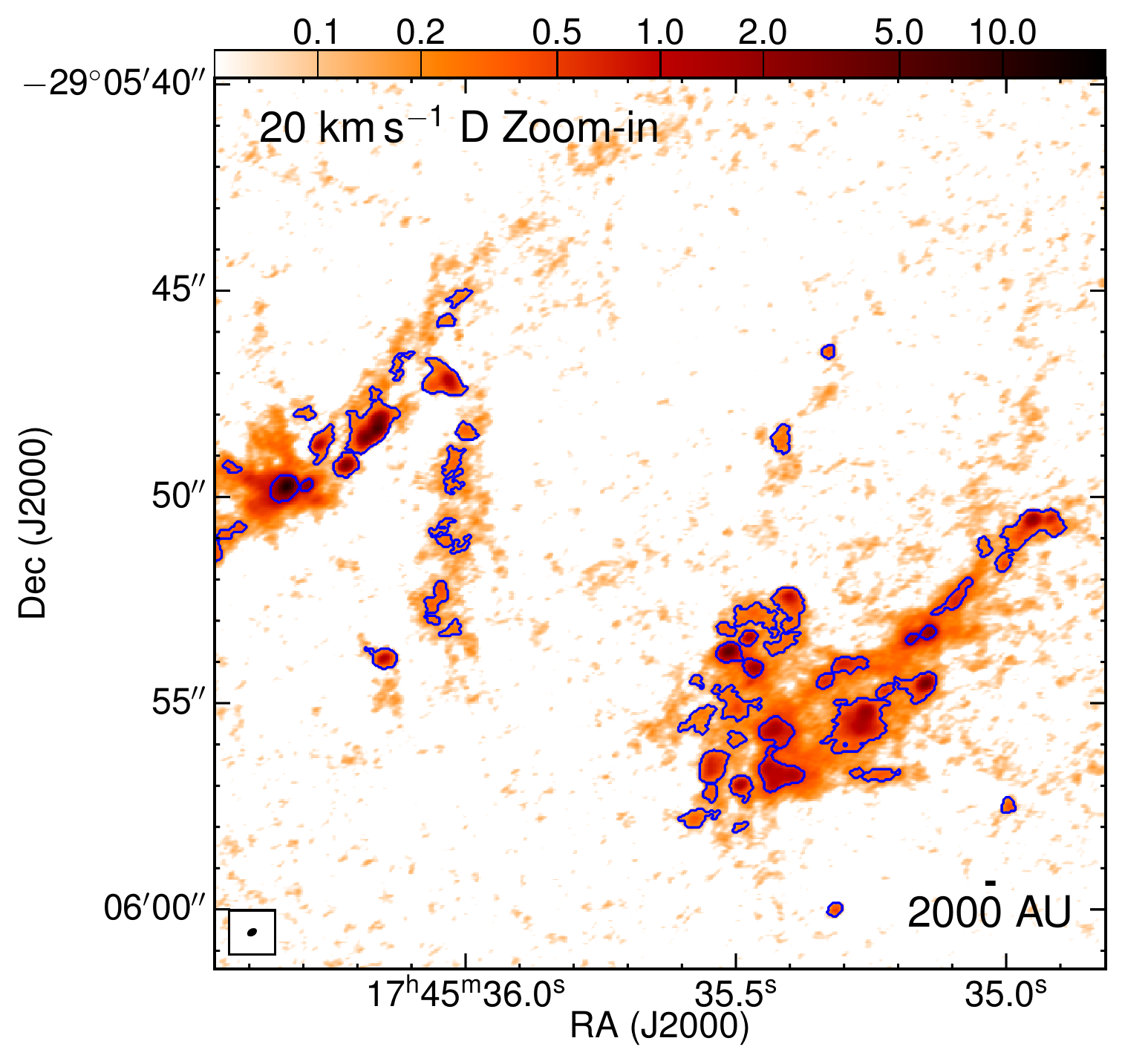} \\
\includegraphics[width=0.45\textwidth]{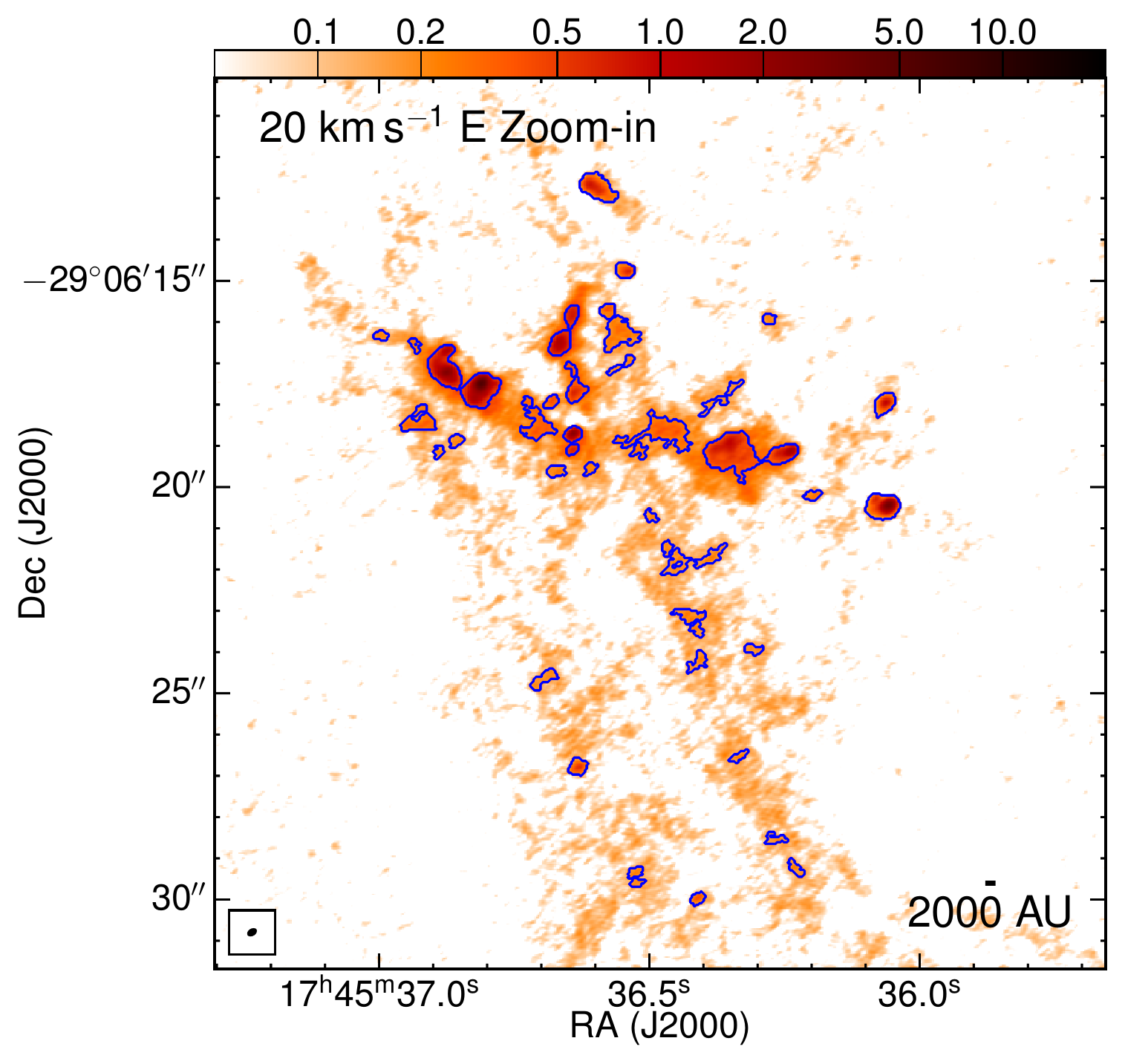} &
\includegraphics[width=0.45\textwidth]{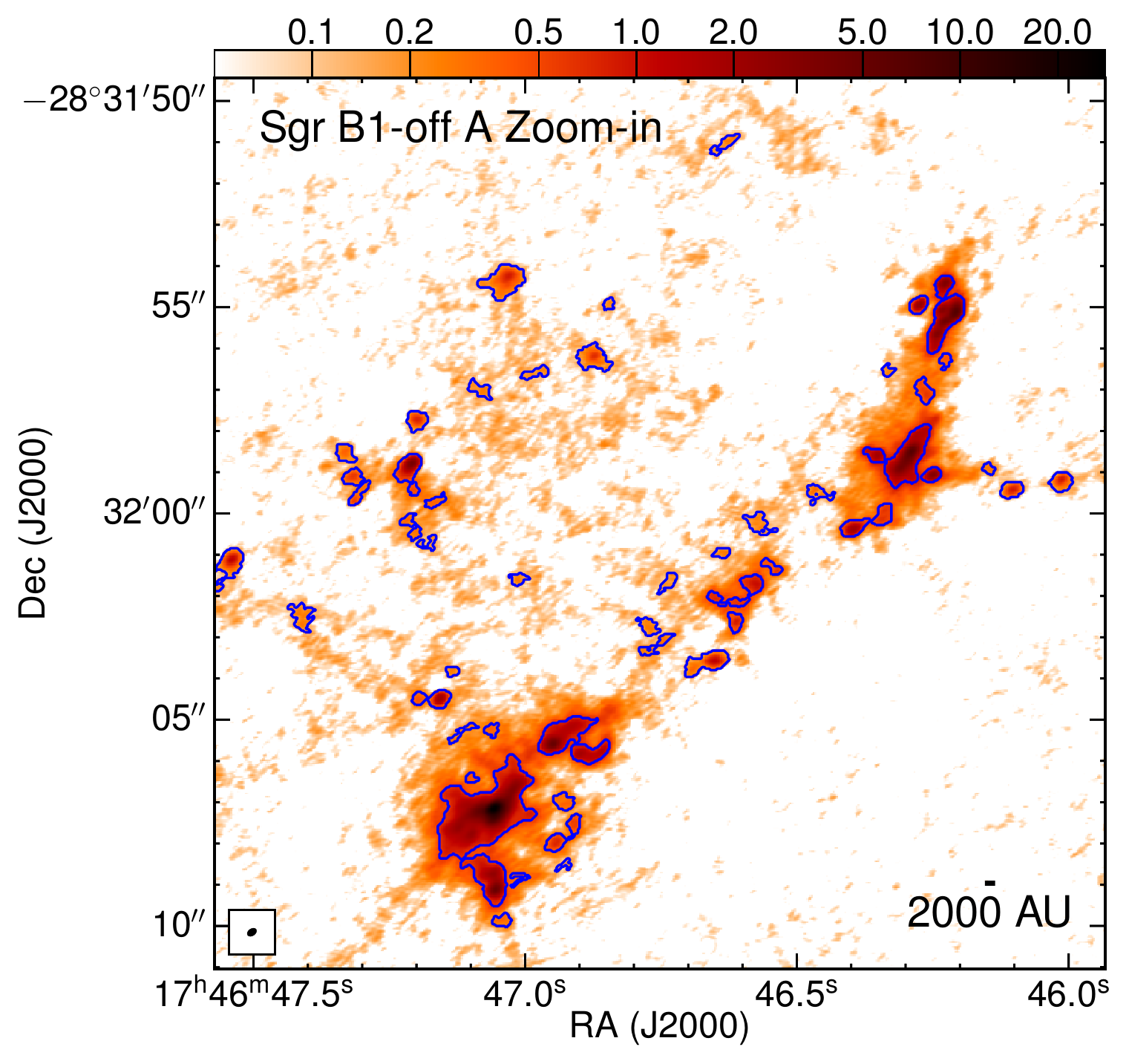}
\end{tabular}
\caption{Zoom-in views of the green boxes in \autoref{fig:dendro}. Blue contours mark the identified cores.}
\label{appd_fig:zooma}
\end{figure*}

\addtocounter{figure}{-1}

\begin{figure*}[!t]
\centering
\begin{tabular}{@{}p{0.45\textwidth}@{}p{0.45\textwidth}@{}}
\includegraphics[width=0.45\textwidth]{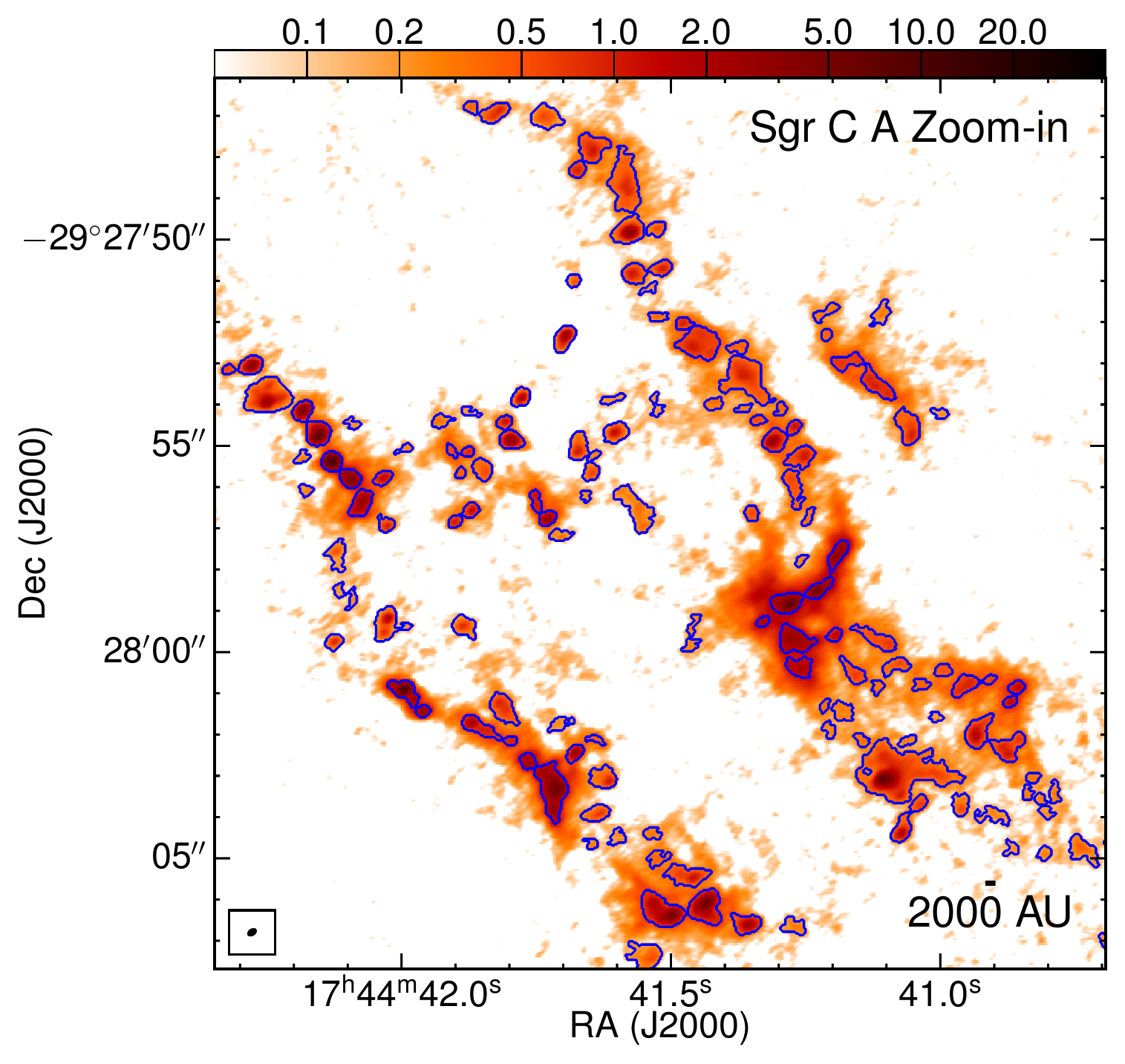} &
\includegraphics[width=0.45\textwidth]{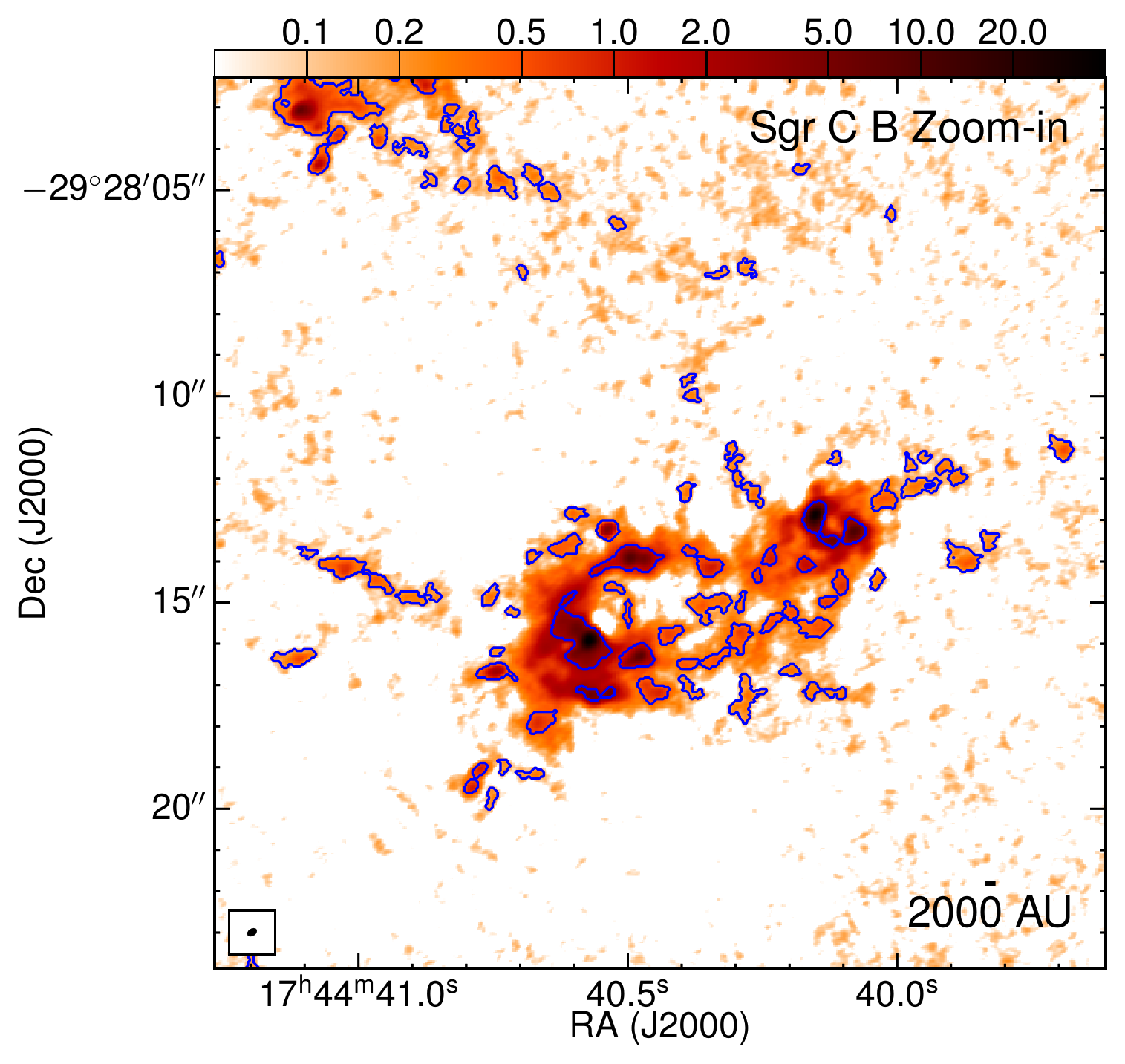}
\end{tabular}
\caption{(\textit{Continued}.)}
\end{figure*}

\section{The Full Core Catalog}\label{sec:appd_catalog}
The full core catalog identified by \textit{astrodendro} is available as a machine-readable table online. The first five entries are shown in \autoref{appd_tab:catalog} as an example.

\begin{deluxetable}{cccccc}
\tabletypesize{\scriptsize}
\tablecaption{The full core catalog in the three clouds.\label{appd_tab:catalog}}
\tablewidth{0pt}
\tablehead{
\colhead{ID} & R.A.~and Decl. & Radius & Flux & Mass & Density \\
 & (J2000) & (AU) & (mJy) & (\msol{}) & (\cc{}) }
\startdata
1 &   17:45:36.41, $-$29:06:30.01 & 1290 & 0.47  & 0.71 & 9.9$\times$$10^6$ \\
2 &  17:45:36.53, $-$29:06:29.63 & 1040 & 0.24 & 0.36 & 9.7$\times$$10^6$ \\
3 &  17:45:36.53, $-$29:06:29.38 & 1240 & 0.34 & 0.51 & 8.2$\times$$10^6$ \\
4 &  17:45:36.23, $-$29:06:29.28 & 1330 & 0.36 & 0.54 & 7.0$\times$$10^6$ \\
5 &  17:45:36.27, $-$29:06:28.57  & 1360 & 0.35 & 0.52 & 6.3$\times$$10^6$ \\
\enddata
\end{deluxetable}

\section{Impact of Varying Dendrogram Parameters}\label{sec:appd_dendro}
We test the impact of different dendrogram parameters on the identified cores (\autoref{subsec:results_fragmentation}), the CMFs (\autoref{subsec:results_properties}), and the MST analysis (\autoref{subsec:disc_jeans}).

\begin{figure*}[!h]
\centering
\begin{tabular}{@{}p{0.45\textwidth}@{}p{0.45\textwidth}@{}}
\includegraphics[width=0.45\textwidth]{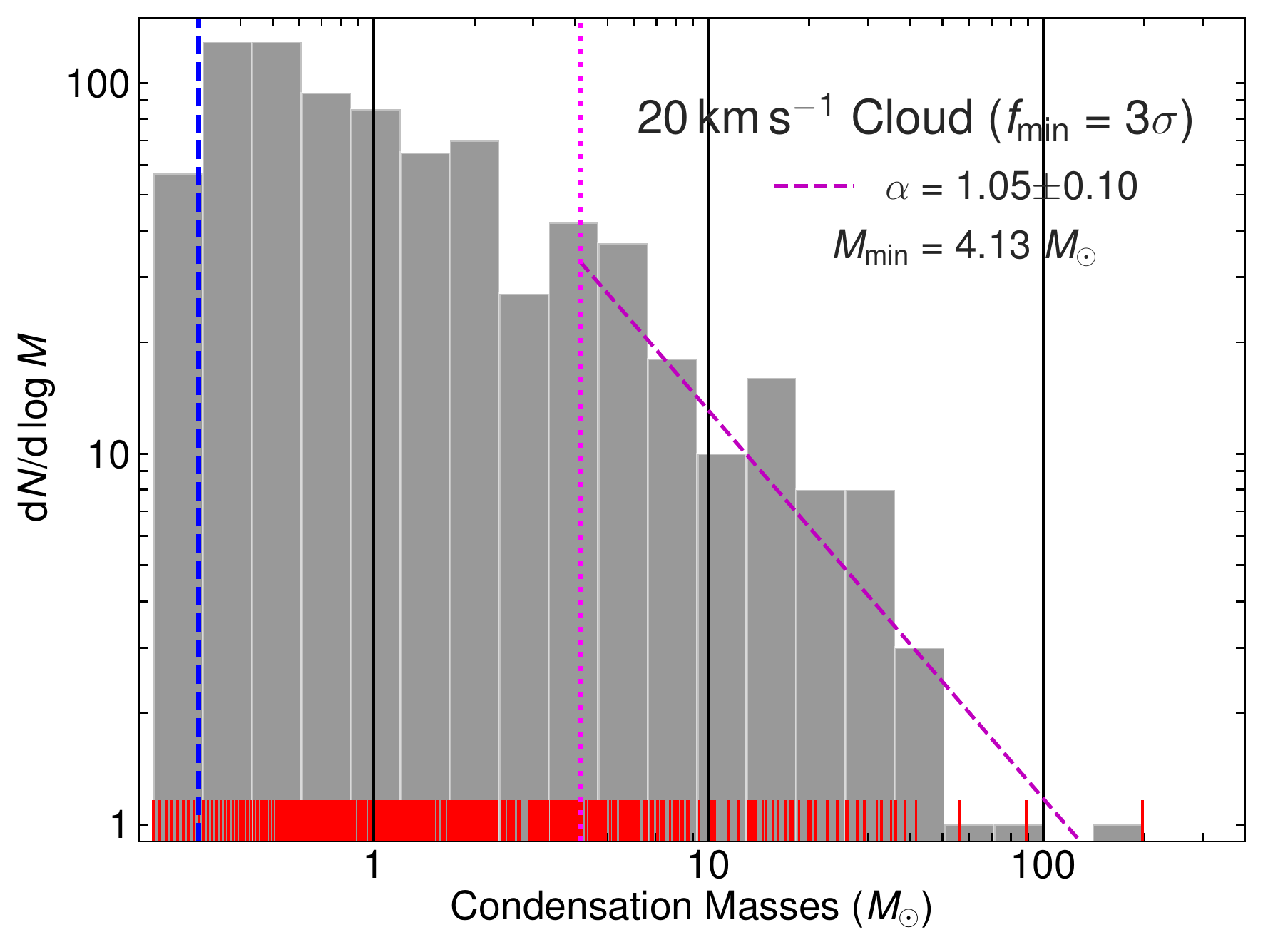} &
\includegraphics[width=0.45\textwidth]{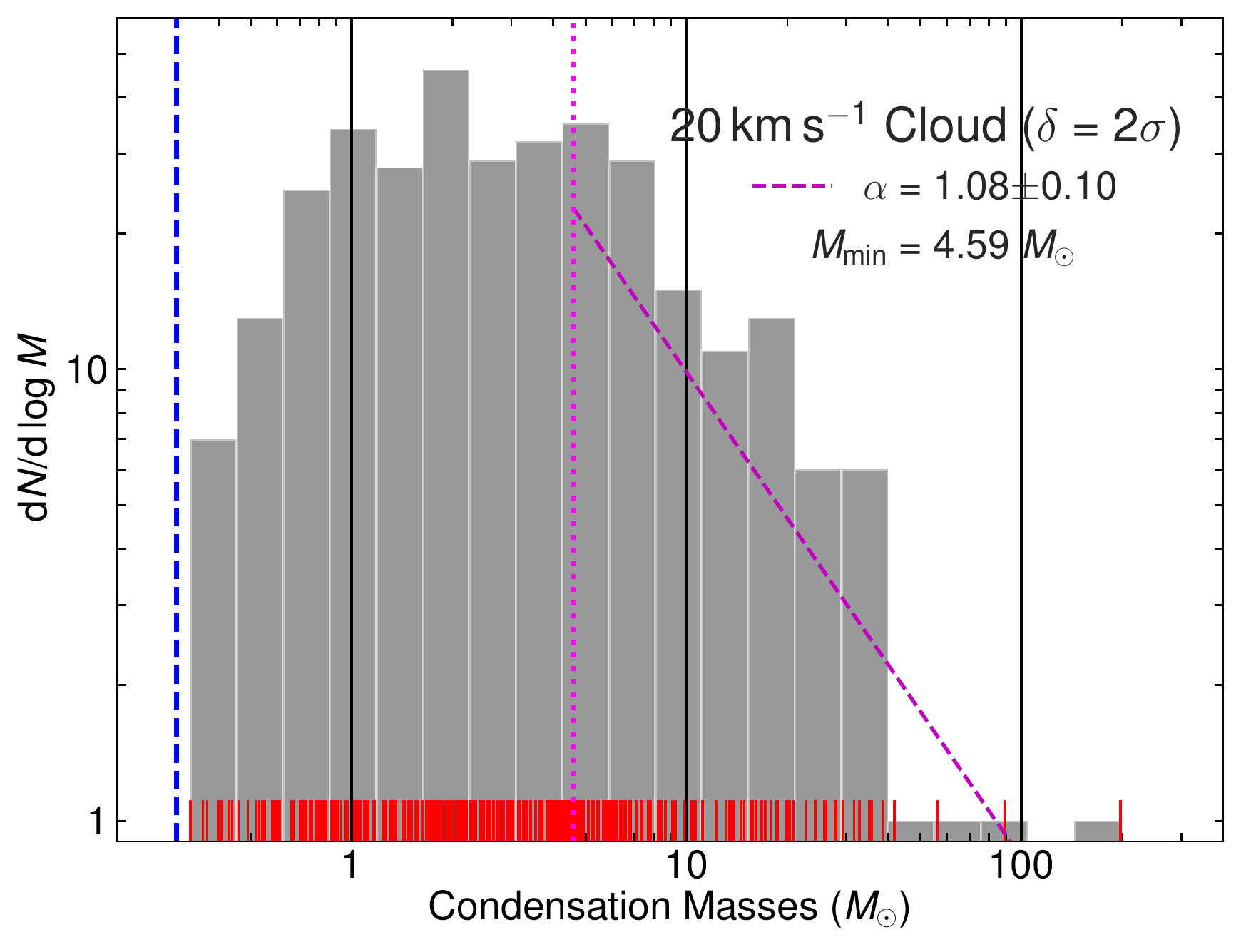}
\end{tabular}
\caption{CMFs of the 20~\kms{} cloud, in the cases of changing the minimum flux density to 3$\sigma$ and the minimum significance for structures to 2$\sigma$. Symbols are the same as in \autoref{fig:cmf}.}\label{appd_fig:cmf}
\end{figure*}

\begin{figure*}[!h]
\centering
\begin{tabular}{@{}p{0.45\textwidth}@{}p{0.45\textwidth}@{}}
\includegraphics[width=0.45\textwidth]{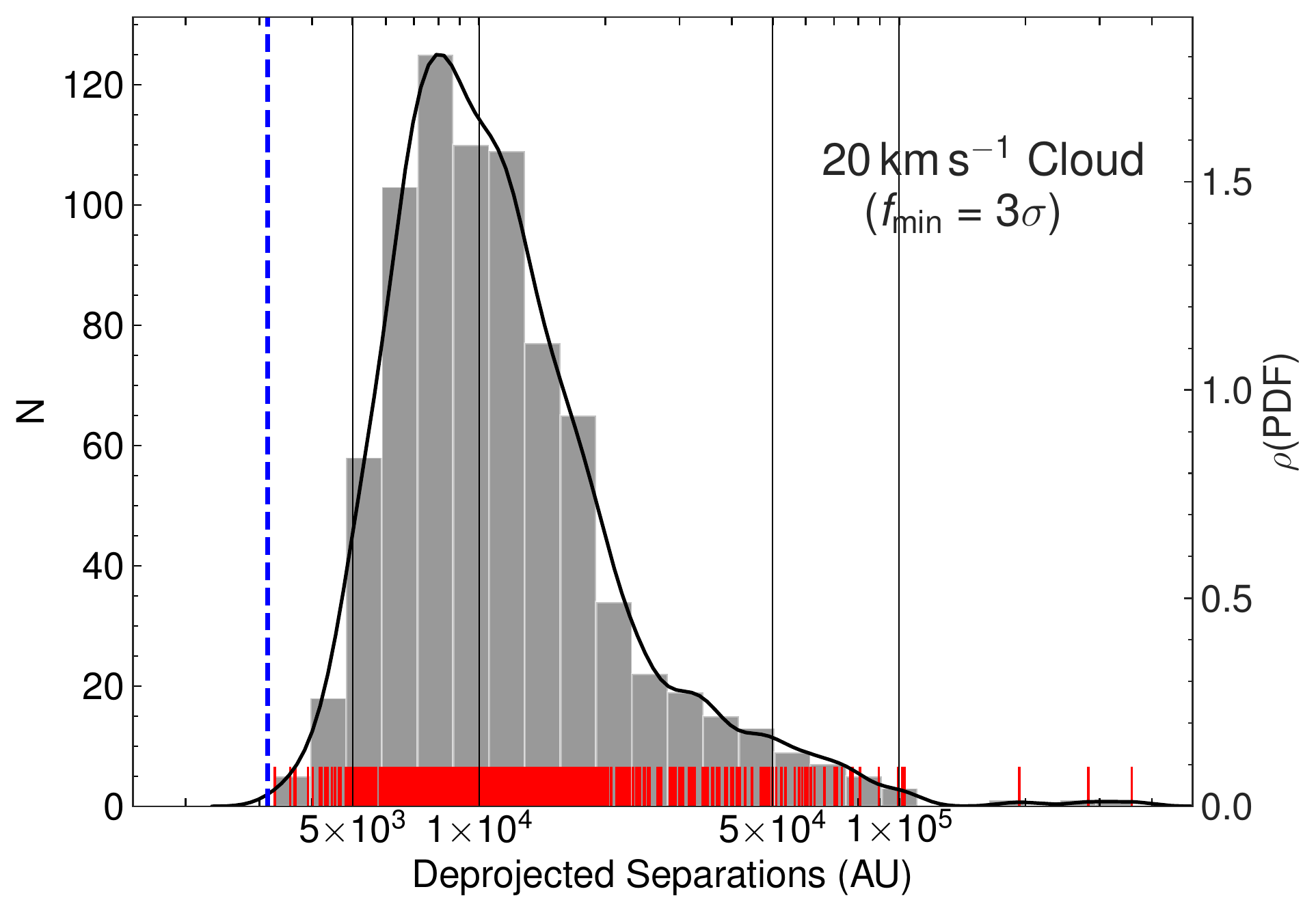} &
\includegraphics[width=0.45\textwidth]{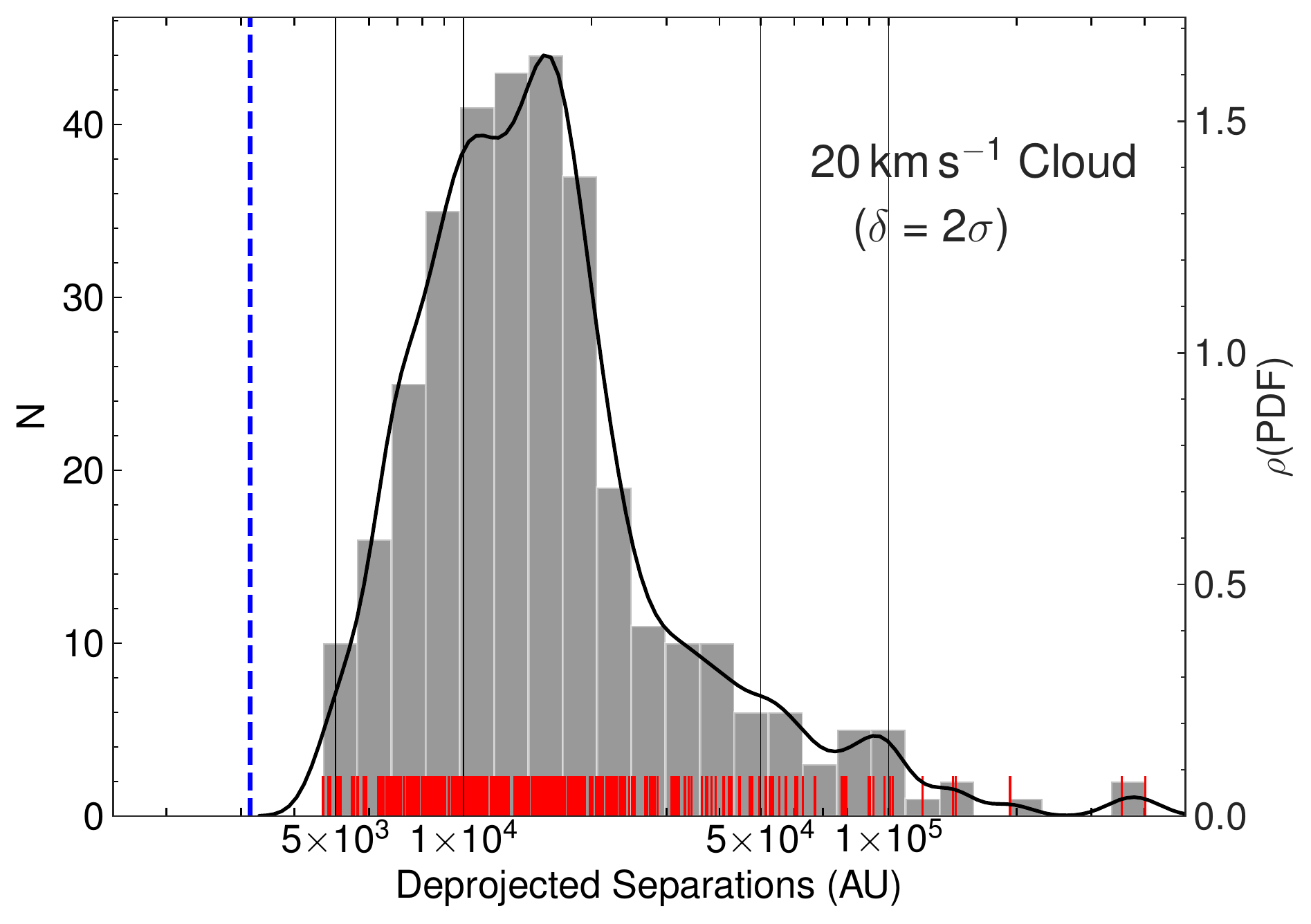}
\end{tabular}
\caption{Distributions of spatial separations between the cores in the 20~\kms{} cloud, in the cases of changing the minimum flux density to 3$\sigma$ and the minimum significance for structures to 2$\sigma$. Symbols are the same as in \autoref{fig:separations}.}\label{appd_fig:mst}
\end{figure*}

First, we change the minimum flux density from 4$\sigma$ to 3$\sigma$. This will identify fainter structures as cores. In the 20~\kms{} cloud, 801 cores are identified (cf.\ 471 cores in the fiducial case). Assuming $T_\text{dust}$=20~K, the smallest core mass is 0.22~\msol{}. The power-law index of the CMF is fit to be $1.05\pm0.10$ starting at the lower bound of 4.13~\msol{} (\autoref{appd_fig:cmf}). The MST analysis yields a characteristic spatial separation of $\sim$8000~AU (\autoref{appd_fig:mst}). 

Second, we change the minimum significance for structures from 1$\sigma$ to 2$\sigma$. This will require any structures to be brighter than the background to be considered as cores. In the 20~\kms{} cloud, 333 cores are identified. Assuming $T_\text{dust}$=20~K, the smallest core mass is 0.33~\msol{}. The power-law index of the CMF is fit to be $1.08\pm0.10$ starting at the lower bound of 4.59~\msol{} (\autoref{appd_fig:cmf}). The MST analysis yields a characteristic spatial separation of $\sim$15000~AU (\autoref{appd_fig:mst}). 

It is possible to vary the the minimum flux density and the minimum significance even further, but then the identification becomes more questionable (e.g., with larger minimum significances, we miss apparent structures; with smaller minimum flux densities, we include spurious detections).

We run the same tests for Sgr~B1-off and Sgr~C, and find consistent results with the 20~\kms{} cloud. In summary, when the dendrogram parameters are changed, the characteristic spatial separations between cores derived from the MST analysis could vary by up to 50\%. The impact on the power-law index of the CMFs is minimal, because different dendrogram parameters mostly affect identification of less bright cores. 

\section{Impact of Varying Dust Temperatures}\label{sec:appd_tdust}
There are two possible biases in the adopted dust temperature of 20~K. First, if we assume that the gas temperature of 50--200~K at 0.1~pc scales continues to smaller scales, and thermodynamic equilibrium between gas and dust at sub-0.1~pc scales, then the dust temperature in the 2000 AU-scale cores would be $\gtrsim$50~K. If we adopt a dust temperature of 50~K, the core masses would decrease by a factor of 3. This does not affect the power-law fitting of the CMFs, though, as all the core masses would decrease synchronously.

Second, as discussed in \autoref{subsec:results_properties}, we cannot rule out the possibility that many of the cores are protostellar thus internally heated by protostars. This may not be a severe issue for low-mass protostars as they usually do not raise dust temperatures higher than 20~K at 2000~AU scales \citep{launhardt2013}, but may be the case for intermediate and high-mass protostars that can heat surrounding dust up to 100--200~K \citep{longmore2011}. For example, all the four cores with $>$100~\msol{} under the current assumptions turn out to be associated with known UC \hii{} regions \citep{lu2019a,lu2019b}, where gas temperatures of $>$150~K have been measured \citep{walker2018}. To account for this, we assume $T_\text{dust}=150$~K for the brightest cores and 20~K for the dimmest ones, and interpolate assuming a power-law dependence of dust temperatures on core fluxes, following \citet{sadaghiani2020}. For the 20~\kms{} cloud, the dust temperature is formulated as
\begin{equation}\label{appd_equ:tdust}
T_\text{dust}=32.8\left(\frac{F_\nu}{1~\text{mJy}}\right)^{0.31}~\text{K}.
\end{equation}
Then the most massive core is 20.6~\msol{} (cf.\ 198~\msol{} with $T_\text{dust}$=20~K), and the second most massive core is only 12.0~\msol{}. Note that at high temperatures the dust opacity $\kappa_\nu$ in \autoref{equ:coremass} may be larger than the adopted value \citep{ossenkopf1994}, so the masses may be overestimated.

With the MLE method in \autoref{subsec:results_properties}, the power-law index of the CMF is fit to be $2.14\pm0.96$ starting at the lower bound of 6.87~\msol{}. However, with this lower bound, only 5 cores are considered in the fitting. Therefore, we also attempt to fix the lower bound to 1~\msol{} to include 180 cores into consideration, and fit a power-law index of $1.40\pm0.10$. In either case, the power law becomes much steeper.

Alternatively, the dust temperatures may not have a well-defined dependence on the continuum fluxes, but rather be randomly distributed. We randomly assign dust temperatures between 20 and 150~K to the 471 cores in the 20~\kms{} cloud and fit the power-law index of the CMFs. By repeating the procedure 1000 times, we find a mean power-law index of 1.34 with a standard deviation of 0.36, which is steeper than the fiducial case for the 20~\kms{} cloud.

The results for Sgr~B1-off and Sgr~C are similar: when assuming higher dust temperatures for brighter cores or random dust temperatures for all cores, the power law in the CMFs significantly steepens and is no longer top-heavy as compared to the IMF. We stress that the above result is by no means physically meaningful, but only serves as an illustration of the impact of dust temperatures on core masses. For a more comprehensive discussion of the uncertainties involved in the CMFs, readers are referred to \citet{cheng2018} and \citet{sadaghiani2020}.

\section{Interpretation of the Lower Bound in the Power-law Fitting to the CMFs}\label{sec:appd_lowerbound}
In \autoref{subsec:results_properties}, we used the MLE method to fit a power law to the high-mass end of the CMF, and at the same time estimated a lower bound. This lower bound corresponds to the the location where as many data as possible are included in the fitting while the data beyond it approach a power law as much as possible. It is chosen by minimizing the distance (represented by a Kolmogorov–Smirnov statistic) between the probability distribution of the measured data and the best-fit power-law model \citep{clauset2009}.

The lower bound can be explained by the confusion limit that leads to inefficient identification of lower-mass cores in clustered environments. To demonstrate this, we used the method in \citet{cheng2018} to estimate the completeness limit of the core sample, by generating artificial cores of certain masses, randomly putting them in the original image, and using \textit{astrodendro} to check whether they can be picked out. If the artificial cores are restricted to be within the 30\% primary beam response (i.e., the same criterion with which we identify cores in \autoref{subsec:results_fragmentation}), the 90\% completeness limit is estimated to be $\sim$2~\msol{} for the 20~\kms{} cloud. If the artificial cores are restricted to a higher threshold, e.g., above a 5$\sigma$ level in the original image, which means they preferably show up in clustered environments, the 90\% completeness limit increases to $\sim$5~\msol{} for the 20~\kms{} cloud. This suggests that in the clustered regions, cores of $\lesssim$5~\msol{} are significantly missed due to the strong background emission, which is likely related to the change of the CMF shape and results in the best-fit lower bound of $\sim$5~\msol{}.

In summary, assuming all the core masses are drawn from a power law distribution, the lower bound can be explained by the confusion limit in clustered regions. There might also be real physical causes (e.g., a real turnover in the CMF), but we cannot confirm them given the aforementioned observational biases.

%\bibliographystyle{aasjournal}
%\bibliography{my}

\end{CJK}
\end{document}